\documentclass[aps,reprint,superscriptaddress,nofootinbib,showpacs]{revtex4-1}
\usepackage{amsmath,latexsym,amssymb,hyperref,graphicx, slashed}

\usepackage{mathrsfs}

\usepackage[T1]{fontenc}

\hypersetup{colorlinks,citecolor= nicegreen,linkcolor= nicered}
\usepackage{color}
\usepackage{hyperref}
\definecolor{red}{rgb}{1,0,0}
\definecolor{nicered}{rgb}{0.7,0.1,0.1}
\definecolor{nicegreen}{rgb}{0.1,0.5,0.1}

\newcommand{\dd}{\mathrm{d}}

\newcommand\GeV{\text{GeV}}
\newcommand\TeV{\text{TeV}}

\newcommand\BB{\ensuremath{0\nu2\beta}}
\newcommand\SEC[1]{\medskip\noindent{\sl\bfseries #1}}

\newcommand{\be}{\begin{equation}}
\newcommand{\ee}{\end{equation}}
\newcommand{\bea}{\begin{eqnarray}}
\newcommand{\eea}{\end{eqnarray}}

\begin{document}


\title{Keung-Senjanovi\'c process at LHC: from LNV to displaced vertices to invisible decays}

\author{Miha Nemev\v{s}ek}
\email{miha.nemevsek@ijs.si}
\affiliation{Jo\v zef Stefan Institute, Jamova 39, Ljubljana, Slovenia}

\author{Fabrizio Nesti}
\email{fabrizio.nesti@units.it}
\affiliation{Dipartimento di Scienze Fisiche e Chimiche, Universit\`a dell'Aquila, via Vetoio SNC, I-67100, L'Aquila, Italy}
\affiliation{INFN Sez.\ Trieste, Via A. Valerio 2, 34127 Trieste, Italy}
\affiliation{Ru\dj er Bo\v{s}kovi\'c Institute, Division of Theoretical Physics, Bijeni\v{c}ka cesta 54, 10000, Zagreb, Croatia}

\author{Goran Popara}
\email{gpopara@irb.hr}
\affiliation{Ru\dj er Bo\v{s}kovi\'c Institute, Division of Theoretical Physics, Bijeni\v{c}ka cesta 54, 10000, Zagreb, Croatia}

\date{\today}

\vspace{1cm}

\begin{abstract} \noindent 
  In the context of Left-Right symmetry, we revisit the Keung-Senjanovi\'c production of
  right-handed $W_R$ bosons and heavy neutrinos $N$ at high energy colliders. We develop a
  multi-binned sensitivity measure and use it to estimate the sensitivity for the entire range of
  $N$ masses, spanning the standard and merged prompt signals, displaced vertices and the invisible
  $N$ region.  The estimated sensitivity of the LHC with 300/fb integrated luminosity ranges from 5
  to beyond 7\,TeV, while the future 33(100)\,TeV collider's reach with 3/ab extends to 12(26)\,TeV.
\end{abstract}



\maketitle

\noindent 
The Standard Model (SM) of fundamental interactions continues to be experimentally verified, and yet
we are short of having evidence for a mechanism providing mass to neutrinos. At the same time, the
weak interactions are evidently parity asymmetric while the fermion sector appears to hint to a
fundamentally parity symmetric spectrum. The Left-Right symmetric theories~\cite{lr, lrspont,
  minkowskims} address these issues simultaneously. The minimal model (LRSM) postulates that parity
is broken spontaneously~\cite{lrspont} along with the breaking of the new right-handed (RH) weak
interaction $SU(2)_R$. The breaking generates at the same time a Majorana mass for the RH neutrino
$N$ and thus also implies Majorana masses of the known light neutrinos via the celebrated see-saw
mechanism~\cite{minkowskims, seesaw}.

Although the scale of breaking is not predicted, the Large Hadron Collider (LHC) would be especially fit for probing this scenario, if the
mass of the new RH gauge boson $W_R$ were in the TeV range. Low energy processes, in particular quark flavor transitions were since
the early times the main reason for a lower bound on the LR scale in the TeV region~\cite{Beall:1981ze, Ecker:1985vv, Mohapatra:1983ae, 
Zhang:2007fn, Maiezza:2010ic, Bertolini:2012pu}. Updated studies of bounds from $K$ and $B$ oscillations~\cite{Bertolini:2014sua} 
and CP-odd $\varepsilon$, $\varepsilon'$~\cite{Bertolini:2012pu} set a lower limit of $M_{W_{R}} \gtrsim 3\text{--}4 \text{ TeV}$, depending
on the measure of perturbativity~\cite{Maiezza:2016bzp, Maiezza:2016ybz} and barring the issue of strong $\mathcal {P}$ 
conservation~\cite{Maiezza:2014ala}. The bottom line is, there remains a significant potential to discover the $W_R$ 
at the LHC or future colliders, with the high scale hinted by tensions in the kaon sector~\cite{Cirigliano:2016yhc}.

\begin{figure}
  \centerline{\includegraphics[width=0.96\columnwidth]{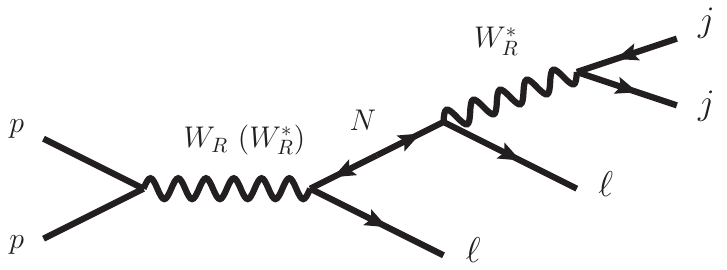}}%
  \caption{The Keung-Senjanovi\'c process. The final state leptons could be of same sign owing to the Majorana nature of $N$. 
  The Drell-Yan production of $W_R$ and $N$ may be dominated by an off-shell $W_R^*$ exchange.}
  \vspace*{-1ex} \label{fig:KS}
\end{figure}

The golden such channel is the Keung-Senjanovi\'c (KS) process~\cite{Keung:1983uu}, in which the Drell-Yan production of $W_R$
generates a lepton and RH neutrino $N$ that in turn decays predominantly through an off-shell $W_R$ into another lepton and two jets,
as depicted in Fig.~\ref{fig:KS}. Due to the Majorana nature of $N$, this process offers the possibility of revealing the breaking of 
lepton number, with the appearance of same sign leptons and two jets.

Pre-LHC studies of the KS process were performed by ATLAS~\cite{Ferrari:2000sp} and
CMS~\cite{Gninenko:2006br}.  Because the heavy neutrino lifetime $l_N$ depends on its mass, the KS
process leads to substantially different signatures depending on $m_N$. A roadmap for different
$m_N$ was performed in~\cite{Nemevsek:2011hz}, using the early LHC data, where transitions from
prompt to displaced and invisible signals were sketched out.

The {\em standard} region is the usual golden channel with $l_N \lesssim 0.02 \text{ mm}$ and two
isolated leptons resulting in the $\ell \ell jj$ signature that was revisited in~\cite{Ng:2015hba,
  Ruiz:2017nip}.  For lighter $N$, it transitions into the {\em merged} region, where one lepton and
two jets merge into a single neutrino (or lepton) jet~\cite{Mitra:2016kov}, the $\ell j_N$
signature. Eventually, the neutrino becomes long-lived and the jet vertex becomes displaced,
$\ell j^d_N$; we call this the {\em displaced} region~\cite{Helo:2013esa, Izaguirre:2015pga, 
Dib:2014iga}. The displaced vertex may lie in the inner detector or even in the external parts
like calorimeters or muon spectrometers. Finally, the {\em invisible} region covers the remaining
case when $l_N \gtrsim 5 \text{ m}$ decays outside of the detector. In this work we systematically
analyze all four relevant regions and provide sensitivity estimates throughout the entire parameter
space.

Existing experimental searches address the standard KS region~\cite{ATLASKS, CMSKS,CMS:2017ilm}, 
while searches for $W'\to \ell \nu$~\cite{ATLASWpMissing, CMSWpMissing} apply to the invisible region. 
However, no active search has been devoted to the merged and displaced regions so far. The purpose of 
this work to provide an assessment of the sensitivity of LHC in these cases and realistically cover the entire
$m_N$ range. We focus our search on $W_R$ masses beyond the limit of $\sim 3.5\,\TeV$, already
excluded by the $W_R\to jj$ search~\cite{ATLASDijet}, and RH neutrino masses
that range from $m_N \sim \text{ few GeV}$, in the invisible region, to $m_N\le M_{W_R}$ beyond
which the process becomes kinematically suppressed.

The $m_N$ region below $\sim 20 \text{ GeV}$ is relevant for phenomenology because of the connection
between the KS process at the LHC and the new physics contributions to neutrinoless double beta decay, 
as studied in~\cite{Mohapatra:1980yp, Tello:2010am, Nemevsek:2011aa}. We will return to this 
interesting connection below.

The work is organized as follows. In the following section we review the kinematics and momentum scales involved
in the KS process, for both on-shell and off-shell $W_R$ production, and describe the diverse resulting signatures. In 
section~\ref{sec:Regions} we study both prompt and displaced regions by simulating the background and signal in order to
assess the sensitivity. In section~\ref{sec:invisible} we study the invisible region where we recast the available search in the 
lepton plus missing energy channel, and also provide the sensitivity at future colliders. Section~\ref{sec:conclusions} 
contains conclusions and an outlook, and a few Appendices contain the analytical details as well as the detailed description
of the binning method used to assess the statistical sensitivity. 

%
%
\section{The Keung Senjanovi\'c process at LHC.}

\noindent 
The minimal LRSM is based on the weak gauge group $\mathcal G_{LR} = SU(2)_L \otimes SU(2)_R \otimes U(1)_{B-L}$ and
a symmetry between the left and right sectors with equal gauge couplings $g_L = g_R$. Correspondingly, quarks and leptons
are arranged in LR symmetric representations, $q_{L,R} = (u,d)_{L,R}$ and $\ell_{L,R} = \left(\nu, e \right)_{L,R}$. The $SU(2)_R$
gauge symmetry is broken spontaneously at some high scale together with the discrete LR symmetry, and the new gauge 
bosons $W_R$, $Z_R$ acquire their masses at that scale. For our purposes it is enough to consider the scale as $M_{W_R}$, 
which, for $g_L \approx g_R$, is already limited to be larger than $\sim 3.5\,\TeV$ by the di-jet searches~\cite{ATLASDijet}. 
This also ensures the smallness of the mixing between left and right gauge bosons, which plays no significant role in 
the rest of the paper.

We are focusing on the search for the $W_R$ gauge boson, which has the following charged-current interactions
\begin{equation}
  \frac{g_R}{\sqrt 2} \left[ V^q_R \, \bar u_R {\slashed W}_R^+ d_R + V^{\ell}_R \,\bar N \slashed W_R^+ \ell_R \right] + {\rm h.c.} 
\end{equation}
where, suppressing flavour indices, $V^q_R$ is the RH analog of the Cabibbo-Kobayashi-Maskawa mixing matrix, 
and $V^{\ell}_R$ is the flavour mixing matrix of RH neutrinos $N \equiv \nu_R$. The RH quark mixing angles inside $V^q_R$
are predicted in the LRSM model to be equal or very near to the standard LH mixings~\cite{Zhang:2007fn, Maiezza:2010ic, 
Maiezza:2014ala, Senjanovic:2014pva}. Potentially small deviations play no significant role at colliders and we use the 
standard CKM matrix for the quark sector. 

With the KS process~\cite{Keung:1983uu}, the LRSM offers a golden search for the new interaction 
mediated by $W_R$ in the presence of $N$. Once $W_R$ is Drell-Yan produced, its decay generates an 
on-shell $N$ that further decays through another off-shell $W_R^*$ into two jets plus a lepton or anti-lepton with equal 
probability, owing to its Majorana nature (see Fig.~\ref{fig:KS}). The whole process is kinematically favored in the 
region $M_{W_R} > m_N$.

In contrast to the quark sector, the leptonic mixing matrix $V^{\ell}_R$ is not predicted by the
model. Instead, its entries can be probed directly at the LHC. The KS process allows to look for
different leptonic flavours in the $\ell\ell j j$ signature~\cite{Das:2012ii,Vasquez:2014mxa}. At
the same time, also channels mediated by the Higgs $h$ or triplet Higgs $\Delta$ can be used to
determine the heavy $N$ Majorana mass matrix.  The Higgs option was dubbed the ``Majorana Higgs''
program, where channels such as $h \to NN$~\cite{Graesser:2007yj, Maiezza:2015lza} and $\Delta \to
NN, h \to \Delta \Delta \to 4 N$~\cite{Nemevsek:2016enw, Dev:2016vle, Dev:2017dui} may be used to
discover lepton number and flavour violation, and to measure the Majorana Yukawa couplings thereby
discovering the spontaneous origin of $N$ masses.

Whichever is the source of information, measuring $V^\ell_R$ is essential to predict neutrino Dirac
masses. Because of the LR parity that is built in the theory, an unambiguous connection between the
Majorana and Dirac masses exists, which is transparent in the $\mathcal C$~\cite{Nemevsek:2012iq}
and slightly less so in the case of $\mathcal P$, see~\cite{Senjanovic:2016vxw}. The connection in
turn predicts the Dirac couplings that can be observed at the LHC and low energies~\cite{Nemevsek:2012iq}.

The right-handed character of $W_R$ may be assessed by analyzing the final states angular
correlations~\cite{Ferrari:2000sp}, as studied in~\cite{Gopalakrishna:2010xm} while invariant mass
variables provide an additional handle for disambiguation~\cite{Dev:2015kca}. In addition, the
extent of the Majorana nature character of \(N\) can be characterized by same versus opposite sign
of dileptons~\cite{Das:2017hmg, Gluza:2015goa}.

Historically, searches~\cite{Ferrari:2000sp, Gninenko:2006br, Nemevsek:2011hz, ATLASKS, CMSKS,
CMS:2017ilm} focused on the on-shell production of $W_R$. The LHC however, especially in the
designed high-luminosity phase, as well as future colliders, have the capability of probing higher
masses for which the production may be dominantly off-shell (see for instance~\cite{Ruiz:2017nip},
where the analysis focuses on heavy to intermediate RH neutrino masses). Thus, in this section we
review the features of the KS process in generality by describing the production of the prompt
charged lepton and $N$ via an on- or off-shell $W_R$, making explicit the distribution of final
states, which play a role in the LHC sensitivity.

\begin{figure}[t]
  \centerline{\includegraphics[width=\columnwidth]{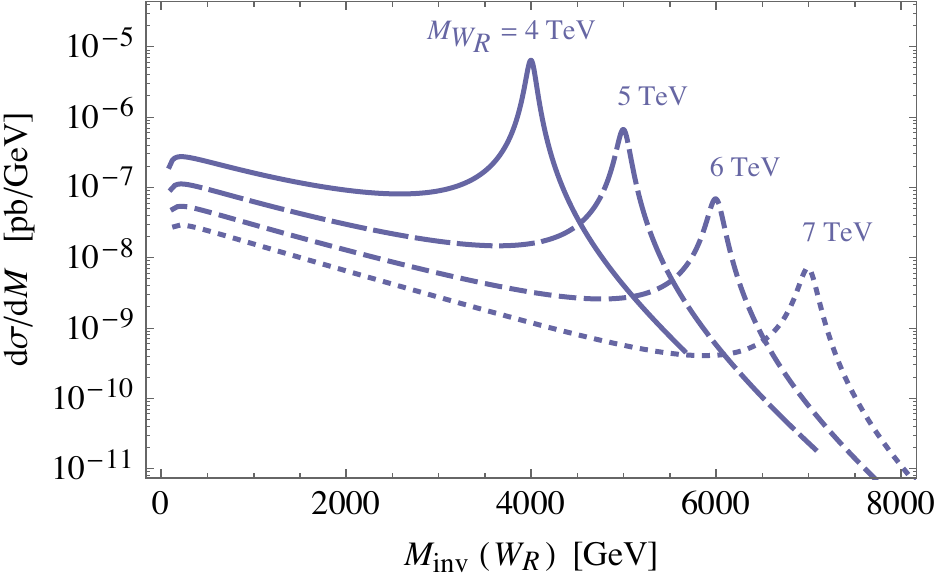}}
  \vspace*{2ex}
  \centerline{\includegraphics[width=\columnwidth]{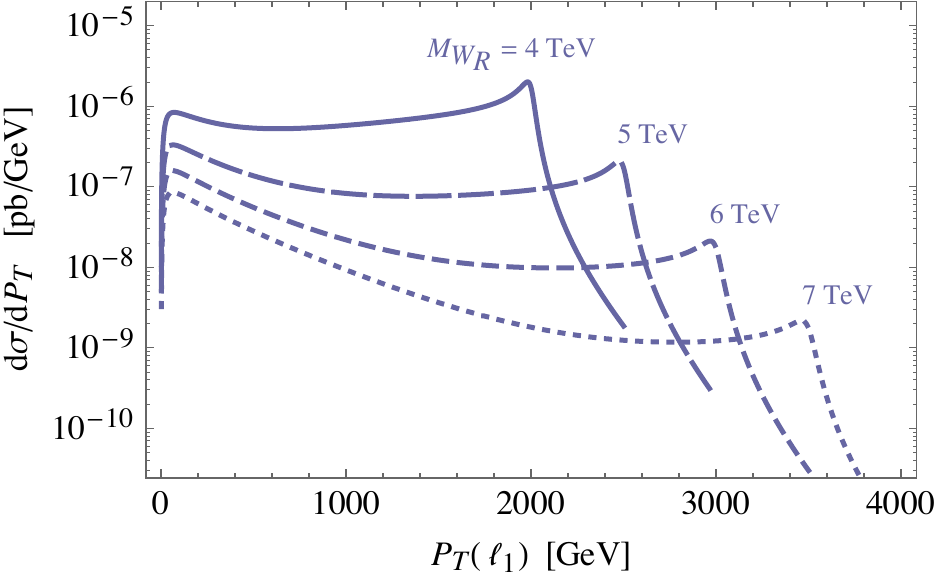}}
  \caption{$W_R$ invariant mass distribution (upper) and primary lepton $p_T$ distribution (lower),
  for $M_{W_R} = 4$--$7\, \TeV$ (solid to dotted). For increasing $M_{W_R}$ the events on
  the on-shell $W_R$ peak become negligible, and the off-shell regime with a low invariant mass
  takes over. Similarly, the primary electron transverse momentum is peaked at low
  values on the lower plot.}
\label{fig:WminvLpt}
\end{figure}

%
%
\SEC{On- and off-shell Drell-Yan production of $W_R$, $N$.}\newline At the LHC the momentum available
from parton constituents is enough to produce an on-shell $W_R$ until $M_{W_R}\lesssim 4\,\TeV$,
with the parton level cross section
\begin{equation}
  \hat{\sigma}_{ij}^{W_R^\pm} (\hat{s}) = \frac{\alpha_2 \pi^2}{3} \left| V^{\text{\tiny CKM}}_{ij} \right|^2
  \delta \left(\hat{s}-M_{W_R}^2 \right)\,.
\end{equation}
For higher $W_R$ masses, the KS process takes place through an off-shell $W_R^*$. Assuming for
simplicity a diagonal coupling of $W_R$ with a single generation lepton and RH neutrino, the parton level
production cross section of $\ell N$ is
\begin{equation}
  \hat{\sigma}_{ij}^{\ell N} (\hat{s}) = \frac{\alpha_2^2 \pi} {72 \hat{s}^2} \left| V^{\text{\tiny CKM}}_{ij} \right|^2
  \frac{\left( \hat{s} - m_N^2 \right)^2 \left( 2 \hat{s} + m_N^2 \right)}{\left( \hat{s}-M_{W_R}^2 \right)^2 + 
  M_{W_R}^2\Gamma_{W_R}^2}
\end{equation}
and we refer to Appendix~\ref{app:offshell} for details.

In the upper plot of Fig.~\ref{fig:WminvLpt} we show the distribution of $W_R$ invariant mass for
the $\ell N$ production at LHC, which shows that the transition between the two regimes is gradual. 
The production clearly becomes dominated by the off-shell contribution when
$M_{W_R} \gtrsim 5\,\TeV$. One sees that the $s$-channel energy involved is always below $\sim \TeV$,
as the $W_R^*$ exchange becomes a contact interaction. A similar effect is seen in the momentum
distribution of $N$ and that of the primary charged lepton $\ell_1$, which is progressively peaked at lower 
energies (lower plot in Fig.~\ref{fig:WminvLpt}). This has implications for the boost inherited by the
neutrino, and thus on its decay length to be analyzed below.

\begin{figure}[t]
  \includegraphics[width=\columnwidth]{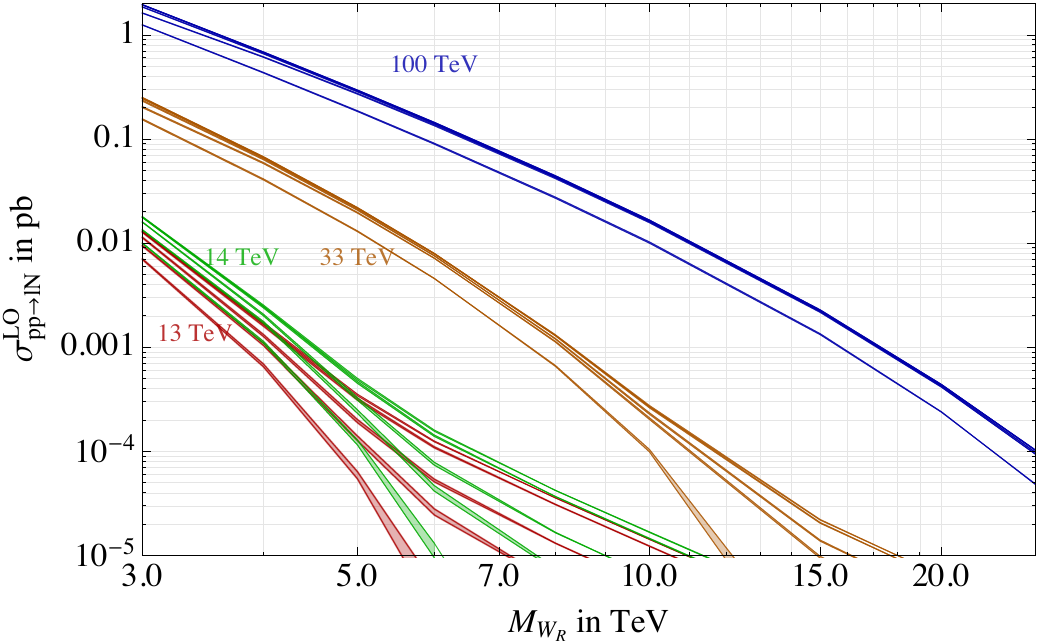}
  \caption{Drell-Yan production cross section of $pp\to W_R^\pm\to \ell^\pm N$. For each indicated 
    interaction energy, the curves from upper to lower are relative to $m_N=50, 100, 500, 1000, M_{W_R}/2$,
    showing normal phase space suppression. In addition, notice the relative enhancement of the lighter $m_N$
    curves for heavier $W_R$, where the $\ell N$ is produced via off-shell intermediate $W_R$. The bands 
    represent the uncertainty due to different PDF sets.}
  \label{fig:WRproduction}
\end{figure}

Taking kinematics and PDFs into account, the $pp \to W_R \to \ell N$ production cross section is shown
in Fig.~\ref{fig:WRproduction} as a function of the $W_R$ mass and for a selection of center of mass
energies and $N$ masses $m_N = \left(50, 100, 500, 1000, M_{W_R}/2 \right) \text{ GeV}$ to
cover both the light $N$ regime up to the standard KS regime.

While heavier $m_N$ are suppressed by phase space, for larger $M_{W_R}$
the off-shell process favors lighter $N$s that show a relative enhancement. Their production
is still significant via $W_R^*$, as long as there is sufficient energy available from the parton 
distribution functions. This has implications for the signals analyzed below.

Indeed, one observes that the regime of light neutrino is particularly promising: already with an
integrated luminosity of $100 \text{ fb}^{-1}$, hadron colliders can probe $W_R$ up to scales
comparable to the available center of mass energy. Keeping in mind the regime of light $N$, we
review the kinematics of its decay at the parton level and as seen by the detector.

%
%
\SEC{Neutrino width and displacement.}
The neutrino width is dominated by the decay into a lepton and a quark pair. Below the top mass, the
width of $N$ is well approximated by
\begin{equation} \label{eq:GN}
  \frac{\alpha_2^2 m_N^5}{64 \pi M_{W_R}^4}
  \sum_{u,c; d,s} \left| V_{ud}^{\text{\tiny CKM}} \right|^2
  \simeq \frac{1}{2.5\,\text{mm}} \frac{(m_N/10\,\GeV)^5}{(M_{W_R}/3\,\TeV)^4}.
\end{equation}
In Appendix~\ref{app:nwidth} we discuss the exact width, valid also for heavier $m_N$.

For progressively lighter $N$ and heavier $W_R$, the lifetime becomes on the order of meters
and in the regime of $m_N\sim10\,\GeV$ the ratio $\Gamma_N/m_N\sim 10^{-12}$ becomes tiny,
leading to issues with event generation, as described in Section~\ref{sec:Regions}.

\begin{figure}[t]
\includegraphics[width=\columnwidth]{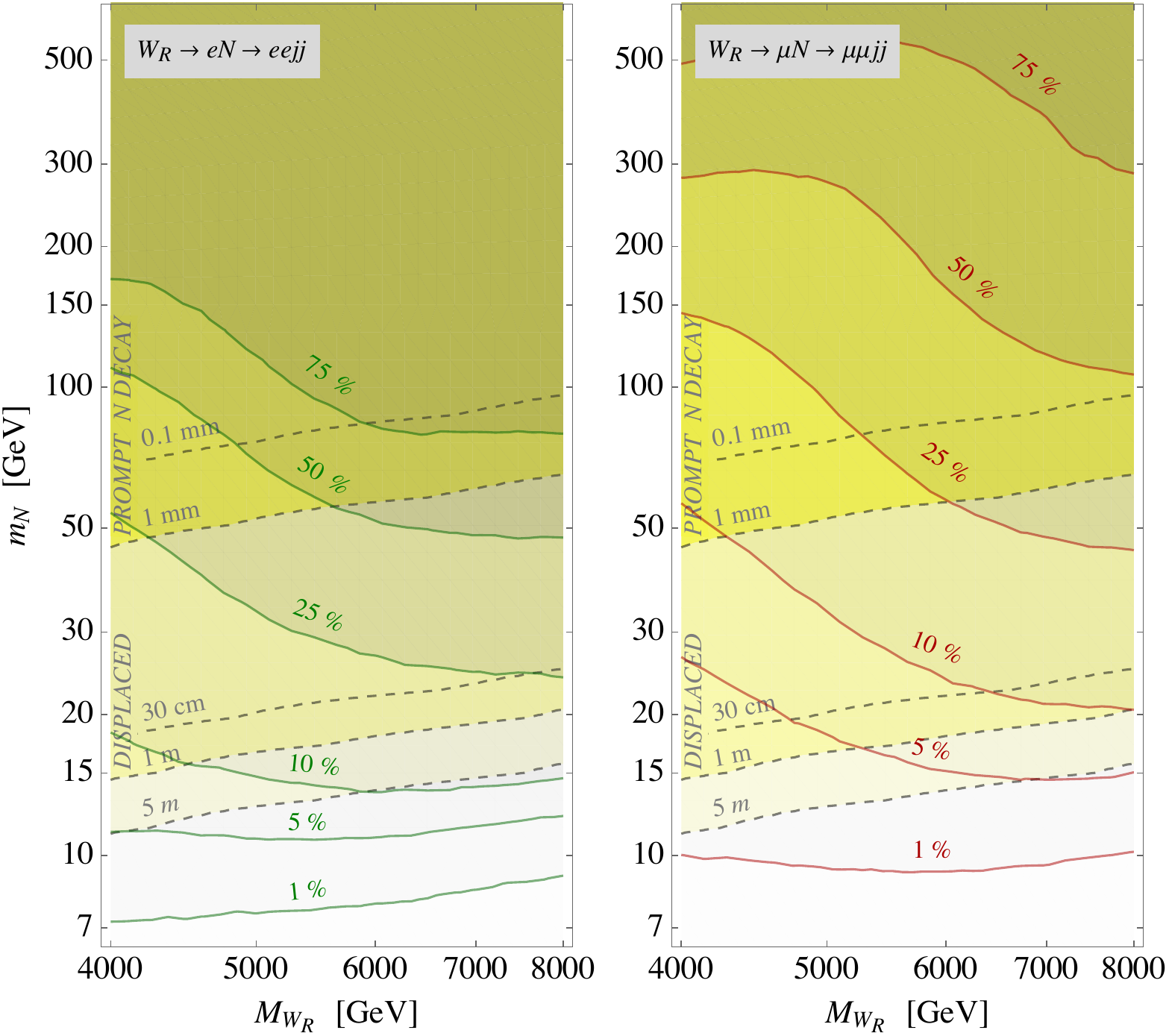}
\caption{Left (right) plot: percentage of secondary leptons passing the isolation requirements 
  is shown by the solid green(red) contours for the electron (muon) channel. Their average displacement
  from $N$ decay is shown in dashed black contours, and the yellow shaded regions mark the prompt
  or displaced $N$ decays showing that $\sim 25 \%$ of electrons and  $\% 10$ of muons are isolated
  below 100 GeV. The lower white region corresponds to decays of $N$ outside of the inner tracker at 
  $\gtrsim 30 \text{ cm}$ or the entire detector $\gtrsim 5 \text{ m}$.}
\label{fig:isolation}
\end{figure}

The decay length in~\eqref{eq:GN} is further increased by the boost from the 
$W_R$ decay. For instance, in case the $W_R$ is produced on shell and practically at rest, the
boost is simply given by $M_{W_R}/2 m_N$. On the other hand, for higher $M_{W_R}$ the 
$W_R^*$ is necessarily off-shell. Its invariant mass is small but it still transmits momentum to the 
primary lepton and to $N$ from the originating partons (see Fig.~\ref{fig:WminvLpt}). For the LHC,
these boost factors can be approximated by
\begin{equation}   \label{onoffshell}
  \gamma_N \simeq 
  \begin{cases}
        \frac{M_{W_R}}{2 m_N},	& W_R \to \text{on-shell}, \\[1ex]
    	\frac{1 \text{ TeV}}{m_N}, 	& W_R \to \text{off-shell},
  \end{cases}
\end{equation}
where the second estimate was performed by the Monte Carlo study. For e.g.\ $m_N=10\,\GeV$ the
boost factor changes from a maximum of $\sim 250$, to the asymptotic $\sim 100$. 
Fig.~\ref{fig:isolation} reports such laboratory decay length including this transition. 

%
%
\SEC{Lepton isolation.} For lower neutrino masses, the boost factor reduces the angular distance
between the secondary lepton and the final jet(s) originating from $N$ decay.  As soon
as this angle goes below the isolation parameters required by the experimental detection, the lepton
is not recognized and gets included in the jet instead.\footnote{The isolation criterion for charged
leptons were imposed by requiring the ratio of charged lepton $p_T$ with respect to the sum of
$p_T$ of surrounding tracks to be larger than 0.15. The adjacent tracks have a threshold $p_T$ of
1 GeV and fit in the cone of $\Delta R = 0.2$ for electrons and 0.3 for muons.}
In Fig.~\ref{fig:isolation} we display the percentage of surviving isolated leptons for the LHC 
at $\sqrt s =  13 \text{ TeV}$. We note that for $M_{W_R} \gtrsim 5 \,\TeV$, where $W_R$ is produced
increasingly off-shell, the $N$ boost declines as in Eq.~\eqref{onoffshell}, such that secondary
leptons are more easily isolated.

Already for $m_N < 70 \text{ GeV}$ where $N$ decays start to become visibly displaced, half or more
of secondary leptons are not isolated anymore. The standard $\ell \ell j j$ case then turns into a
single isolated lepton and another jet containing the secondary lepton, $\ell j$.  The important
conclusion here is that as $m_N$ is lowered, secondary leptons become non-isolated before being
displaced. Thus the secondary lepton will be merged in a completely displaced merged neutrino jet.

In summary, in the light neutrino mass regime, the signature of the process consists typically of a
single prompt lepton and another jet. While this final state does not offer the handle of LNV, it does
show a characteristic displacement of the neutrino jet. Eventually for very low RH neutrino mass, 
the entire displaced jet is generated outside the detector and manifests as missing energy.

To analyze these different signatures, we separate the cases in four regions as outlined in 
the introduction:
\begin{enumerate}
  \item The \emph{standard} KS region, which for LHC requires $m_N\gtrsim $150--200\,\GeV, features 
  two leptons and two jets ($\ell \ell jj$). The leptons are of same sign in half of the cases due to
  the Majorana nature of $N$, and the invariant masses $m_{\ell \ell jj}^{\text{inv}}$ and 
  $m_{\ell jj}^{\text{inv}}$ can reconstruct the masses of $W_R$ and $N$. 
  \item The \emph{merged} region where the signature is a prompt lepton and a jet containing the products
  of $N$ decay including the secondary lepton ($\ell j_N$). The small mass of $N$ makes it difficult
  to reconstruct its mass through the $j_N$ invariant mass. Still, $M_{W_R}$ can be identified via
  the invariant mass of $m_{\ell j_N}^{\text{inv}}$.
  \item The \emph{displaced} region where the merged neutrino jet appears at a visibly displaced distance from
  the primary vertex ($\ell j^d_N$).
  \item The \emph{invisible} region where the jet appears outside the detector and manifests itself as missing
  transverse momentum ($\ell \slashed{E}_T$).
\end{enumerate}
The separation between the above regions is not sharp, a fraction of events leaks
from one region to another and eventually results in overlapping exclusion regions.

%
%
\section{The standard, merged and displaced KS} \label{sec:Regions}

\noindent In this section, we assess the reach of the LHC in the standard, merged and 
displaced regions defined above. 

We first discuss the intricacies of event generation and the procedure for identifying
the jet displacement at the detector level. We then describe the relevant backgrounds and
finally adopt a dedicated statistical procedure for assessing the signal sensitivity, designed to
deal with correlated kinematical variables.

%
%
\SEC{Event Generation.}
Commonly used multi-purpose Monte Carlo event generators such as \textsc{MadGraph} are well suited
to simulate the standard KS region. However, difficulties appear in dealing with extremely narrow
resonances, as is the case in the merged, displaced and invisible regions where $\Gamma/M \simeq
10^{-12}$ or less. The difficulties are related to insufficient numeric precision as well as to
phase space integration coverage (see~\cite{thesis} for a detailed discussion). To avoid these
issues and generate a reliable signal, we developed a custom event generator, made available
on~\cite{lrsite} and described in the Appendix~\ref{app:kseg}. It generates events at parton-level,
including the case of the off-shell $W_R$ as well as light or heavy RH neutrino. The NLO corrections
of $W_R$ production, are taken into account with a $K$-factor that is well approximated by a
constant value of 1.3 (see~\cite{Mitra:2016kov} for a recent computation). Events are finally
hadronized using \textsc{Pythia 6}.

The presence of an energetic primary lepton ensures triggering of the events, and leaves us with
just the problem of identifying the possibly displaced jet.

\begin{figure}[t]
  \centerline{\includegraphics[width=.96\columnwidth]{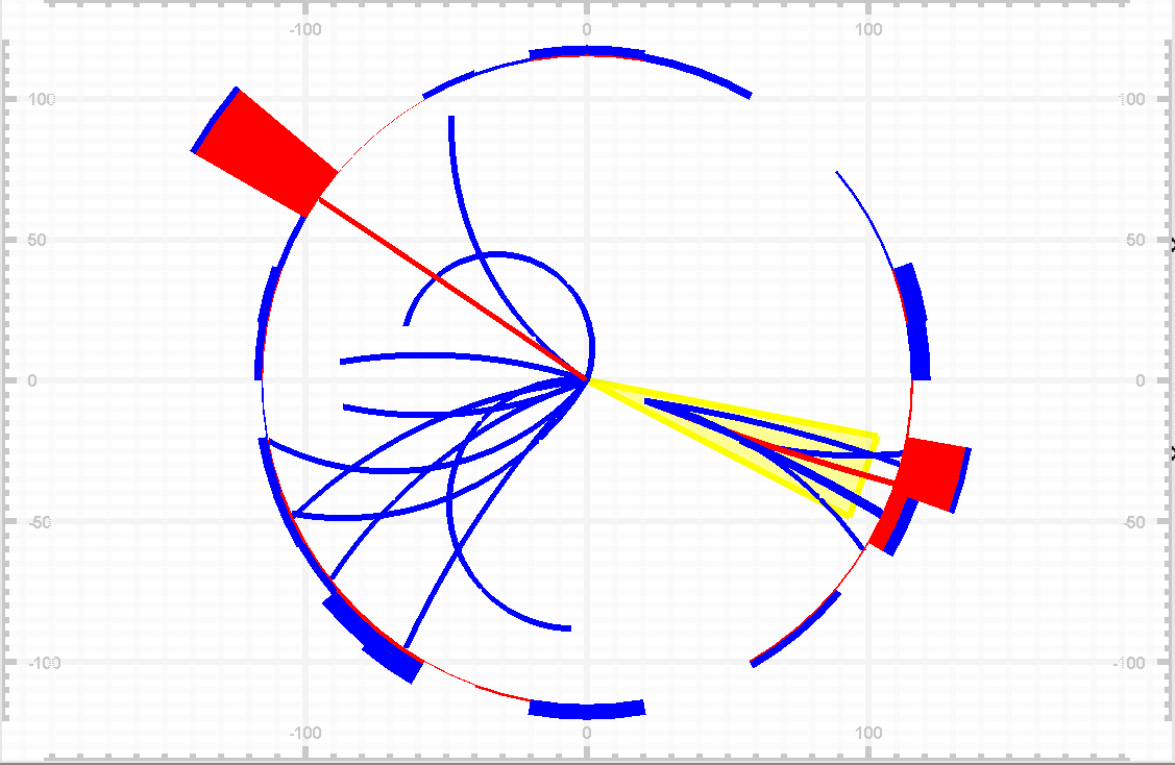}}
  \caption{A typical event featuring the prompt electron and a merged jet on the opposite side,
    including the secondary electron and constituent tracks from a $\sim 3\,$cm-displaced vertex. 
    Both electron tracks are drawn in red. 
    (Picture produced using the \textsc{Delphes}~\cite{deFavereau:2013fsa} event viewer.)
    \label{fig:event}\vspace*{-2ex}}
\end{figure}

%
%
\SEC{Recognizing displaced jets.} At detector level, we have adopted the \textsc{Delphes}
software~\cite{deFavereau:2013fsa}, improved by developing a custom
module for jet displacement recognition. 

The problem of identifying the displacement of the jet origin is quite nontrivial for a number of
reasons, mainly because inside of each jet a number of tracks with displaced origin are typically
always present (due to decays of long-lived hadrons like e.g. $B$-mesons) and make part of the jet
sub-structure. Moreover, a number of soft tracks are coming from the primary vertex processes that
usually accompany any displaced hard process. These make it hard to detect its presence. A number of
approaches to cope with these problems, i.e.\ to probe the jet-substructure have been devised that
suit particular scenarios. The strategy that we adopt is as follows: jet displacement is defined as
the minimum displacement among the tracks associated with the jet which have $p_T$ larger than some
threshold, calibrated to $50\,\GeV$. This simple but robust algorithm reproduces the correct
displacement in 95\% of the signal cases. In Fig.~\ref{fig:event} we display a typical event where
the displaced jet can be recognized by the displaced vertex from which its most energetic
constituent tracks are originating.

It is worth mentioning that in defining each track displacement, also the smearing of the track
vertex position due to (momentum dependent) detector resolution was implemented
~\cite{vertexsmearing}). The minimal resolution is $\sim$0.01--0.02--0.1\,mm, therefore below these
values no displacement can in any case be appreciated.  We do not apply extra suppression factors
due to efficiency of displaced vertex recognition. In this regard, we note that at displacements
between few millimeter and few centimeters, vertex efficiency is typically large $\sim
80\%$~\cite{Aaboud:2017iio}, while a dedicated vertexing algorithm may need to be implemented to
detect displacements below few millimeters. On the other hand, we discard jets with
displacement beyond 30\,cm, for which the vertex reconstruction by tracking appears largely unfeasible.

Finally, momentum resolution is also important especially for muons, because for one it gets
progressively worse for large momentum $\sim \TeV$, and because the secondary muon can become part
of the jet, thus contributing to its invariant mass. As a benchmark, we assume the momentum
resolution as studied in~\cite{momentumresolution} for the ATLAS detector.
\begin{figure*}[t]
\def\www{.695}
\centerline{%
\includegraphics[height=\www\columnwidth]{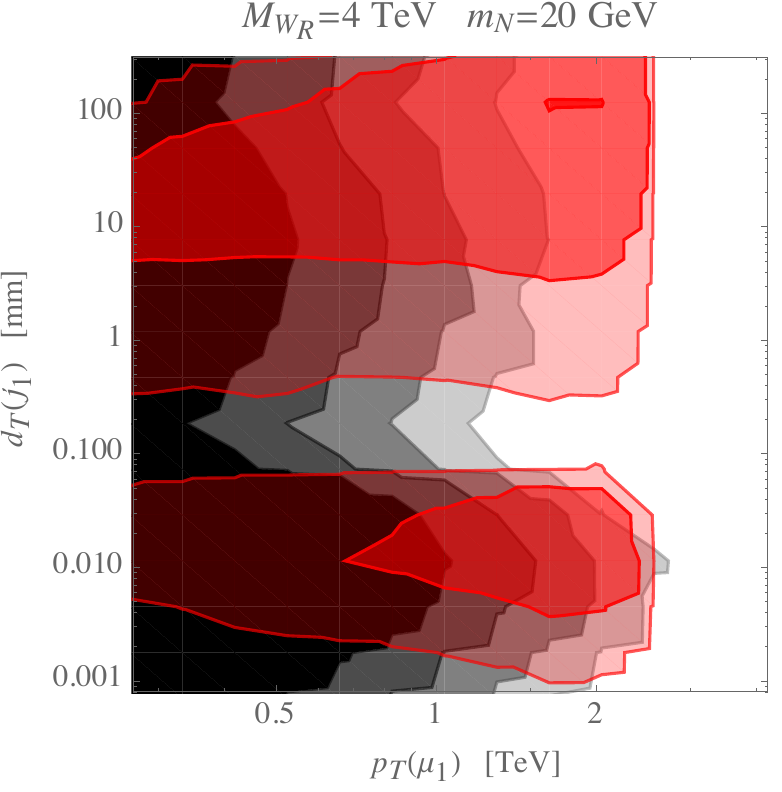}~%
\includegraphics[height=\www\columnwidth]{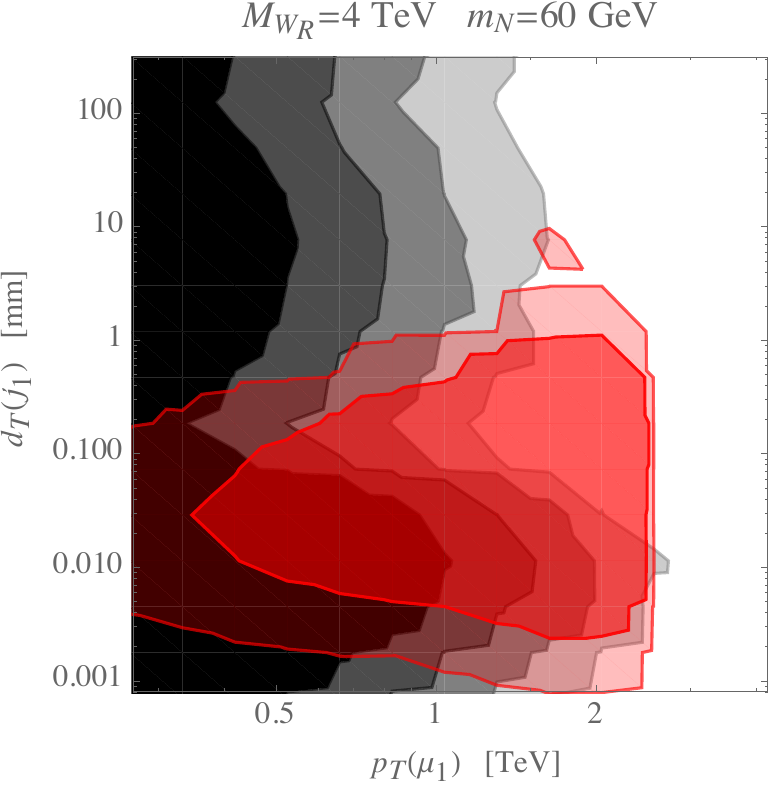}~%
\includegraphics[height=\www\columnwidth]{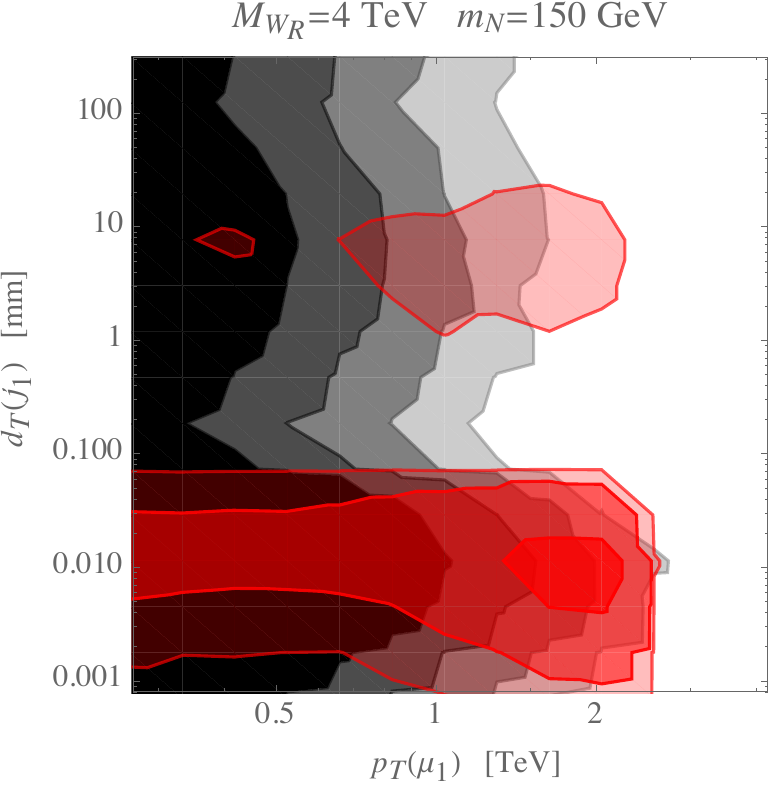}%
}
\vspace*{1ex}
\centerline{%
\includegraphics[height=\www\columnwidth]{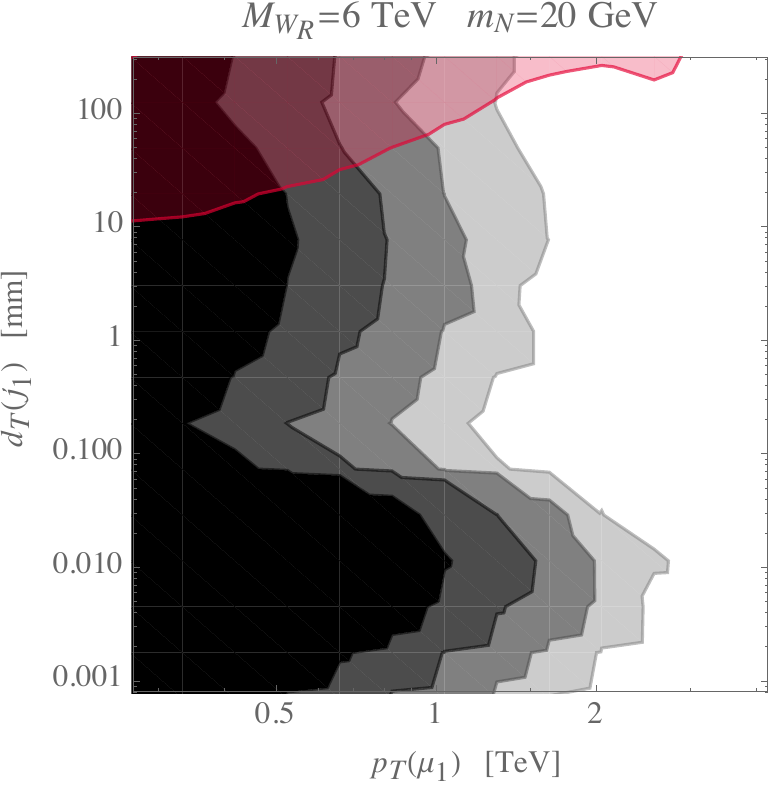}~%
\includegraphics[height=\www\columnwidth]{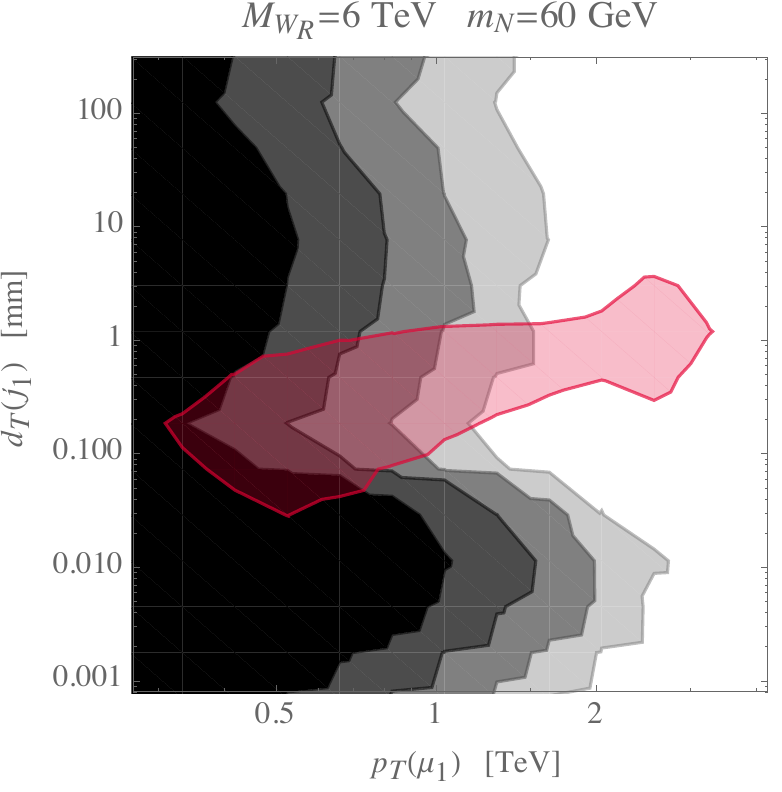}~%
\includegraphics[height=\www\columnwidth]{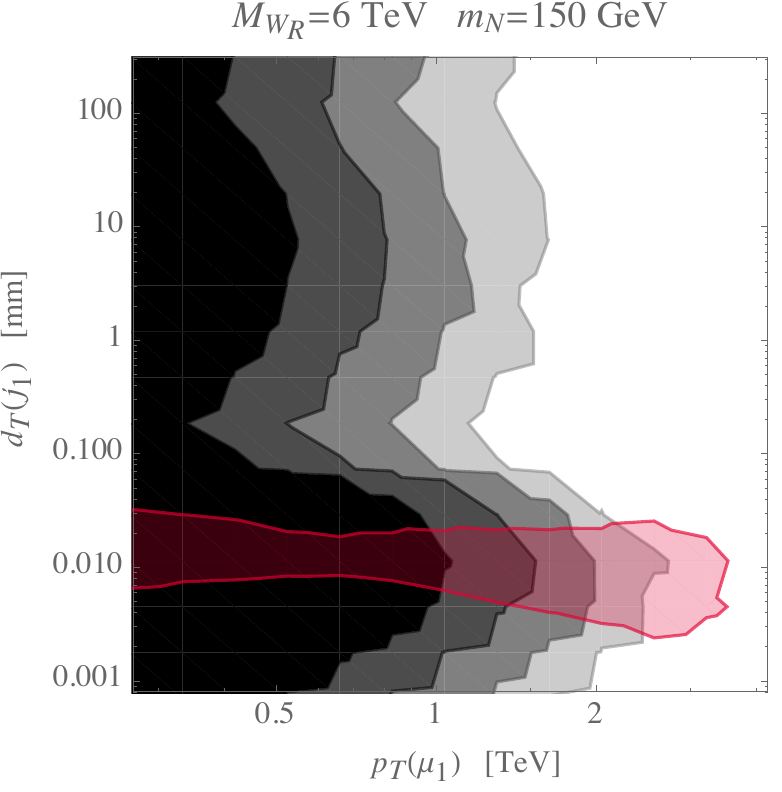}%
}%
\vspace*{-1ex}
\caption{Event distribution in $p_T$ and displacement of the hardest jet. Shown are background
  (gray) and signal (red) for some sample values of $M_{W_R} = 4, 6 \text{ TeV}$ (upper, lower) and
  $m_N = 20, 60, 150 \text{ GeV}$ (left to right).  The distributions are exemplified with a binning
  grid of $15 \times 15$, the increasingly dark shading referring to bins with respectively more
  than $0.1$, $1$, $10$, $100$ events.}\label{fig_pte_vs_dtj}
\end{figure*}

%
%
\SEC{Backgrounds.} The dominant backgrounds contributing to this process are production of single or
double vector bosons plus jets as well as production of $t \bar t$ plus jets.\footnote{Additional
backgrounds from so called jet fakes, i.e.\ jets misidentified as leptons, are found to be
negligible in~\cite{CMS:2017ilm} in the standard KS region; in the merged and displaced regions
its effect can be suppressed by asking tight isolation of the prompt lepton.}

While prohibitive to generate in full strength, we can take advantage of the fact that due to
Eq.~(\ref{onoffshell}) the parton momenta in the signal are very rarely less than a few hundred
GeV. Thus the background can be efficiently generated by imposing a cut of minimal $p_T>150\,\GeV$
at parton level without loosing the signal. We use a stable version \textsc{MadGraph} 2.3.3,
\textsc{Pythia} 6 and modified \textsc{Delphes} 3 with the anti-$k_T$ jet clustering algorithm with $\Delta R=0.3$.  The
number of background events simulated at generator level with the relative weights $\ll1$, as well
as the events recognized at detector level are:

\begin{center}
\begin{tabular}{|l|c|l|c|}
  \hline
  ~~background ~ & ~\# generator~ & ~weight~ & ~\# detector~~\\
  \hline
  ~~$V+012j$ & 22.46\,M& ~~0.021 & 9.93M\\
  ~~$VV+012j$ & 10.55\,M & ~~0.0028& 4.61M\\
  ~~$t \bar t+012j$ &10.47\,M & ~~0.024 & 4.38M\\
\hline
\end{tabular}
\end{center}

These are strongly reduced to respectively 378k, 15.6k, 65k expected detector level events when
restricting the relevant kinematical variables to their loose range of interest (see below the first
column of Table~\ref{tab:flow}).  A basic cut on $p_T(\ell)\gtrsim 1\,\TeV$ could reduce them
further to $\sim$ 250, 20, 7, or even less without sacrificing more than 20\% of signal. Instead of
adopting this rough procedure, we describe in the next paragraph a more efficient method of
assessing the sensitivity.

\begin{figure*} \centerline
  \centerline{\includegraphics[width=0.97\columnwidth]{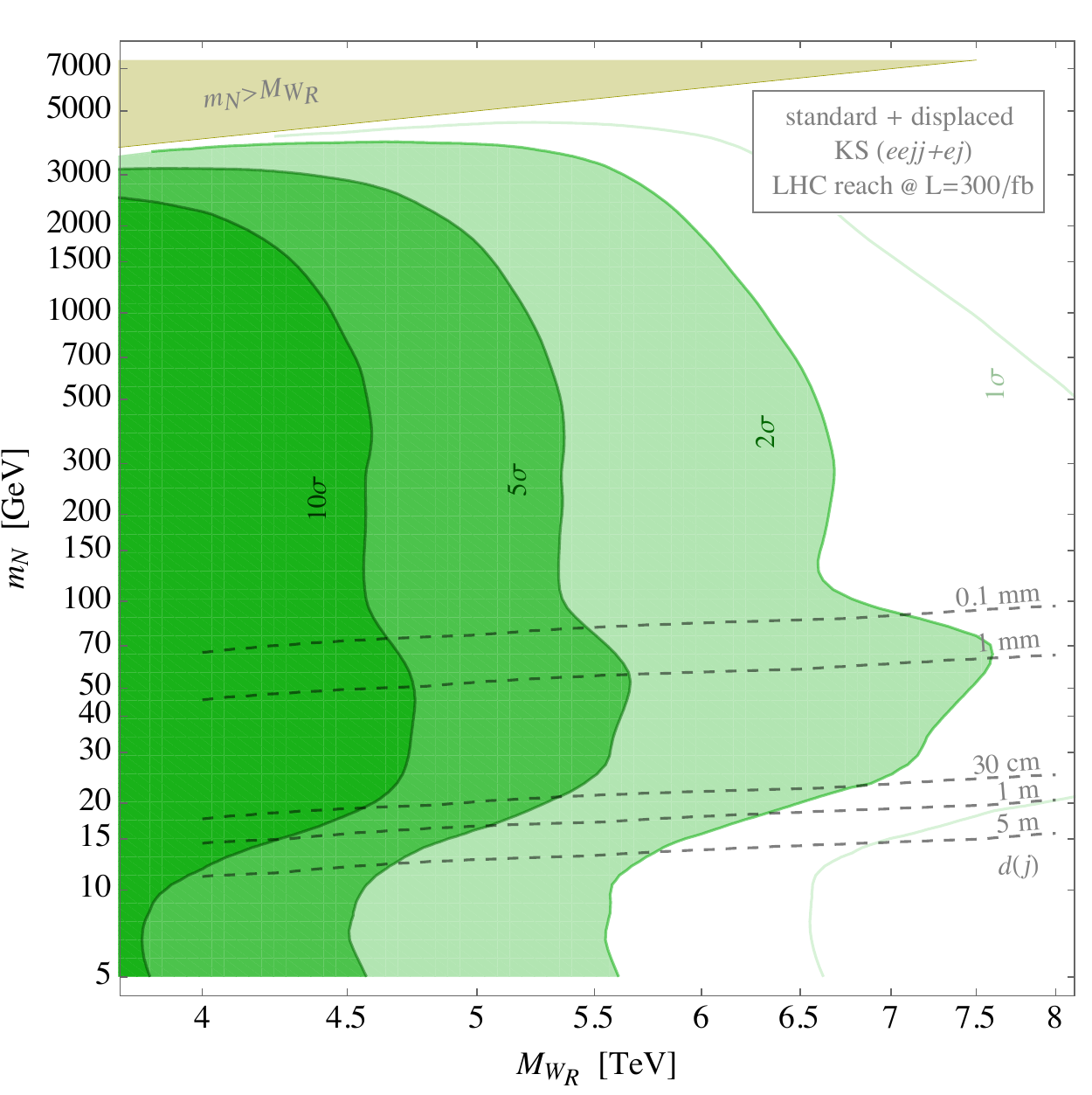}%
~~~~~\includegraphics[width=0.97\columnwidth]{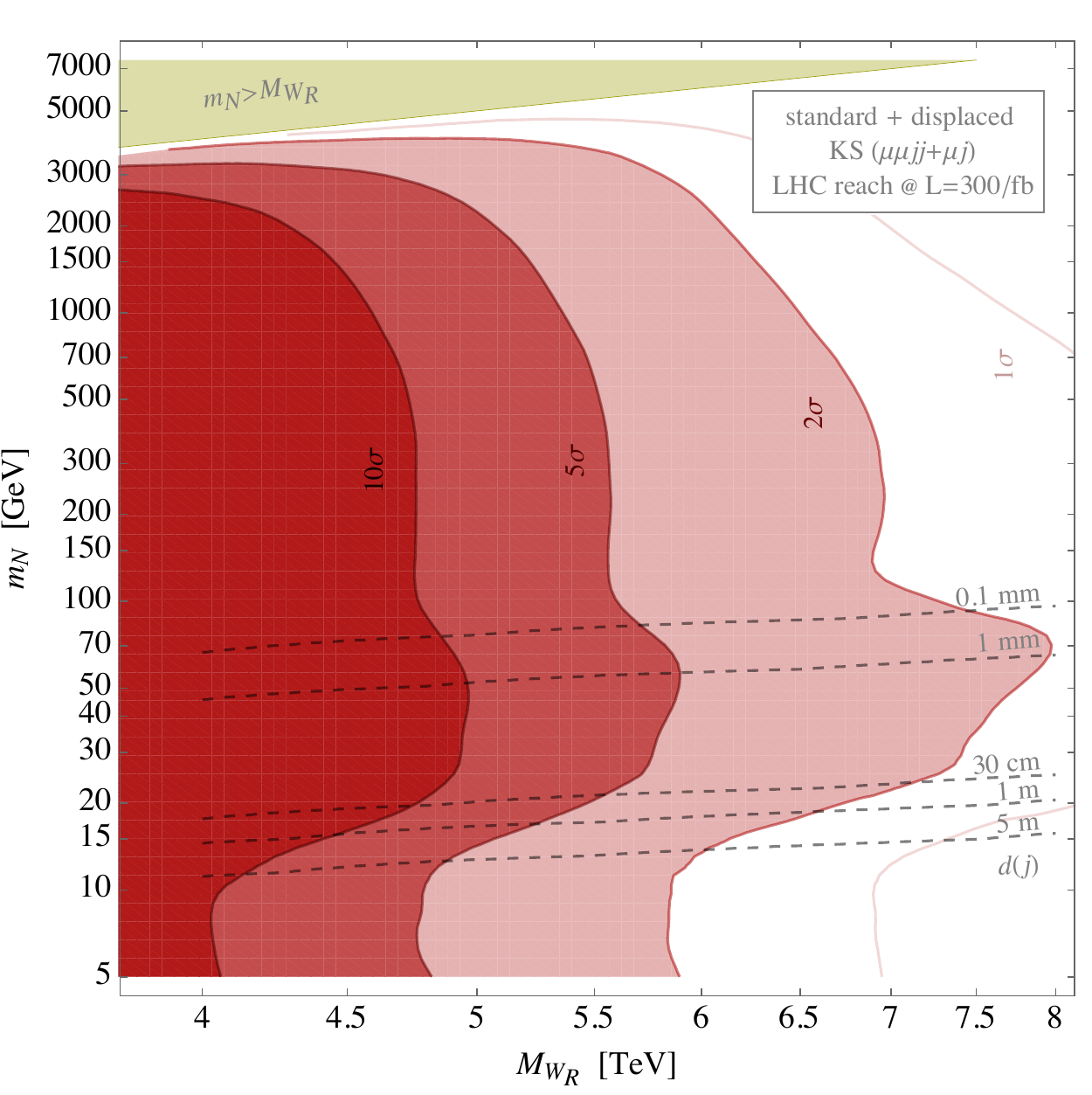}}
  \caption{LHC sensitivity to the KS signal in the $M_{W_R}$ -- $m_N$ plane, for integrated luminosity
  of $\mathcal L=300/$fb. Left, green (Right, red) frames show the sensitivity in the electron
  (muon) channel, obtained by combining the prompt $\ell \ell j j $ signature which features LNV as
  well as the displaced $\ell j j $ signature. Contours show the LHC reach at 1, 2, 5, 10 $\sigma$
  C.L.}
\label{fig_sens_300}
\end{figure*}

\begin{table*}[t]
\vspace*{2ex}
$
\begin{array}{|lcc|rcccccccc|}
\hline
~~~~~~\text{\color{nicegreen} Electron Channel}\!\!\!\!\!\!\!\!\!\!\!\!\!\!\!\!       & ~~~~~~~~~~~~\mathcal{L}=300\,\mathrm{fb}^{-1} \!\!\!\!\!\!\!\!\!\!\!         && M_{W_R}\!: & 4\, \TeV  & 4\,\TeV   & 4\,\TeV &~~ & 6\,\TeV  & 6\,\TeV   & 6\,\TeV & 6\,\TeV \\
\text{variable}              & \text{range}              & \text{\# bins} & m_{N}\!:    & 20\,\GeV  & 300\,\GeV & ~2\,\TeV~ & & 20\,\GeV & 300\,\GeV & ~2\,\TeV~ & ~3\,\TeV~ \\
\hline                                                                                                                                                       
p_T(\ell_1)                  & \{150,4500\}\,\GeV        & {35}                     &   & 14.19     & 13.82     & 7.19    & & 1.03     & 1.77      & 1.22    & 0.80 \\
d_T(j_1)                     & \{0.001, 300\}\,\text{mm} & {100}                    &   & 17.57     & 14.04     & 7.60    & & 2.02     & 1.91      & 1.38    & 0.97 \\
\# (\text{jets})             & 1,2,3,4                   & {4}                      &   & 17.88     & 14.20     & 7.94    & & 2.24     & 2.04      & 1.47    & 1.08 \\
\#(\text{leptons})           &  1,2                      & {2}                      &   & 17.97     & 14.90     & 9.08    & & 2.30     & 2.23      & 1.60    & 1.22 \\
\#(\text{same sign})         &  0,1                      & {2}                      &   & 18.00     & 15.71     & 9.85    & & 2.32     & 2.61      & 1.70    & 1.30 \\
m^{\text{inv}}_{\ell_1 j_1}  & \{200,8500\}\,\GeV        & {20}                     &   & 18.82     & 17.24     & 10.91   & & 2.81     & 3.03      & 1.91    & 1.47 \\[.5ex]
\hline
\end{array}
$
\caption{Grid binning variables and progressive sensitivities obtained with $300\,\text{fb}^{-1}$, for a selection of 
  points in parameter space, representing the regimes of single lepton and displaced jet, single lepton and jet, and
  standard two leptons plus two jets.}
\label{tab:flow}
\end{table*}

%
%
\SEC{Assessing the sensitivity.} Examples of event distributions are reported in
Fig.~\ref{fig_pte_vs_dtj} in the plane of primary lepton momentum versus hardest jet
displacement. We see that as the mass scales vary, the relative position of signal and background
changes. In particular, because the jet displacement for the signal depends strongly on $m_N$, the
signal region can overlap or instead be separated from one or more regions dominated by
backgrounds. In situations like this, the effectiveness of the usual method of devising selection
cuts is limited.

For this reason, instead of adapting the selection cuts to the values of model parameters, we prefer
to devise a simpler and more robust method to assess the sensitivity.  The method is a
straightforward multi-bin generalization of the usual $s/\sqrt{s+b}$ measure relative to single bin
Poisson-counting experiments. It combines single bin sensitivities of a multidimensional grid as
the sum in quadrature, including the bins dominated by backgrounds,
\begin{equation} \label{eq:sensitivity}
  \text{sensitivity}\ \Sigma=\left[\sum_{i\in\text{bins}}\frac{s_i^2}{s_i+b_i}\right]^{1/2}.
\end{equation}
In Appendix~\ref{app:sensitivity} we describe in detail the formal aspects together with
statistical and systematic uncertainties, also commenting on the binning dependence.

The binning grid that we adopt here spans the variables as described in the first column of
Table~\ref{tab:flow}, with broad enough intervals. In choosing the number of bins, we took care not 
to refine the binning below the resolution in the relevant kinematic variable(s). In the same table
we also report the effectiveness of successive binning procedures in different kinematical variables
for a selection of $(M_{W_R},m_N)$. These are representative of the regime of lepton non-isolation
with jet displacement, the standard KS regime with LNV, and also of on-shell versus off-shell $W_R$.

Finally, the maximal statistical and systematic uncertainty on the sensitivity can be quoted as $\pm
0.5$ and $\pm 0.01$, as discussed in Appendix~\ref{app:sensitivity}. 

The result of the analysis is shown in Fig.~\ref{fig_sens_300} for both the muon and electron
channel. Starting from below, i.e.\ from the most displaced region, we see that as soon as the
displacement of neutrino decay can be detected by the tracker, i.e.\ below 30\,cm, displacement
helps in raising the sensitivity, which features a bump, for masses up to $m_N \sim
40$--$60\,\GeV$. Thus, in this region, even if LNV is not observable, a very good sensitivity can be
achieved by discriminating on the jet displacement. The result is a promising reach of more than
7\,\TeV, at 95\%\,C.L..

Just above, in the prompt but merged region with $150\,\GeV \lesssim m_N\lesssim \TeV$, the
sensitivity is lower due to phase space suppression. Nevertheless, as soon as genuine LNV becomes
observable, the presence of same sign leptons acts as a complementary variable. In the standard KS
regime where LNV helps, the combined effect leads to a plateau up to circa $m_N \sim\TeV$ or
500\,\GeV, with sensitivity to circa $M_{W_R}\sim 6.5\,\TeV$ at 95\%\,C.L..

Above that, the KS process becomes increasingly suppressed by kinematics and sensitivity
drops.

\begin{figure*}
  \centerline{%
  \includegraphics[height=0.8\columnwidth]{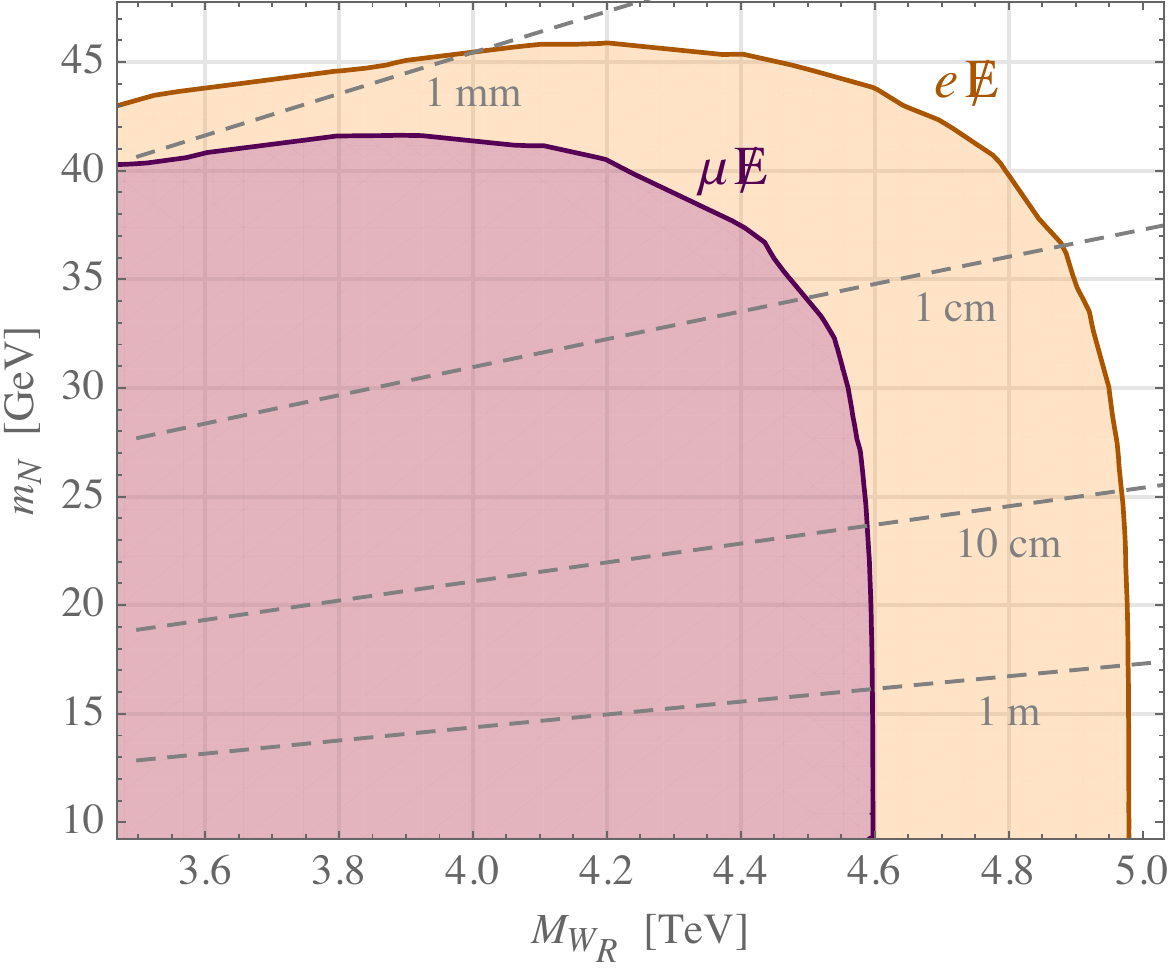}~~
  \includegraphics[height=0.81\columnwidth]{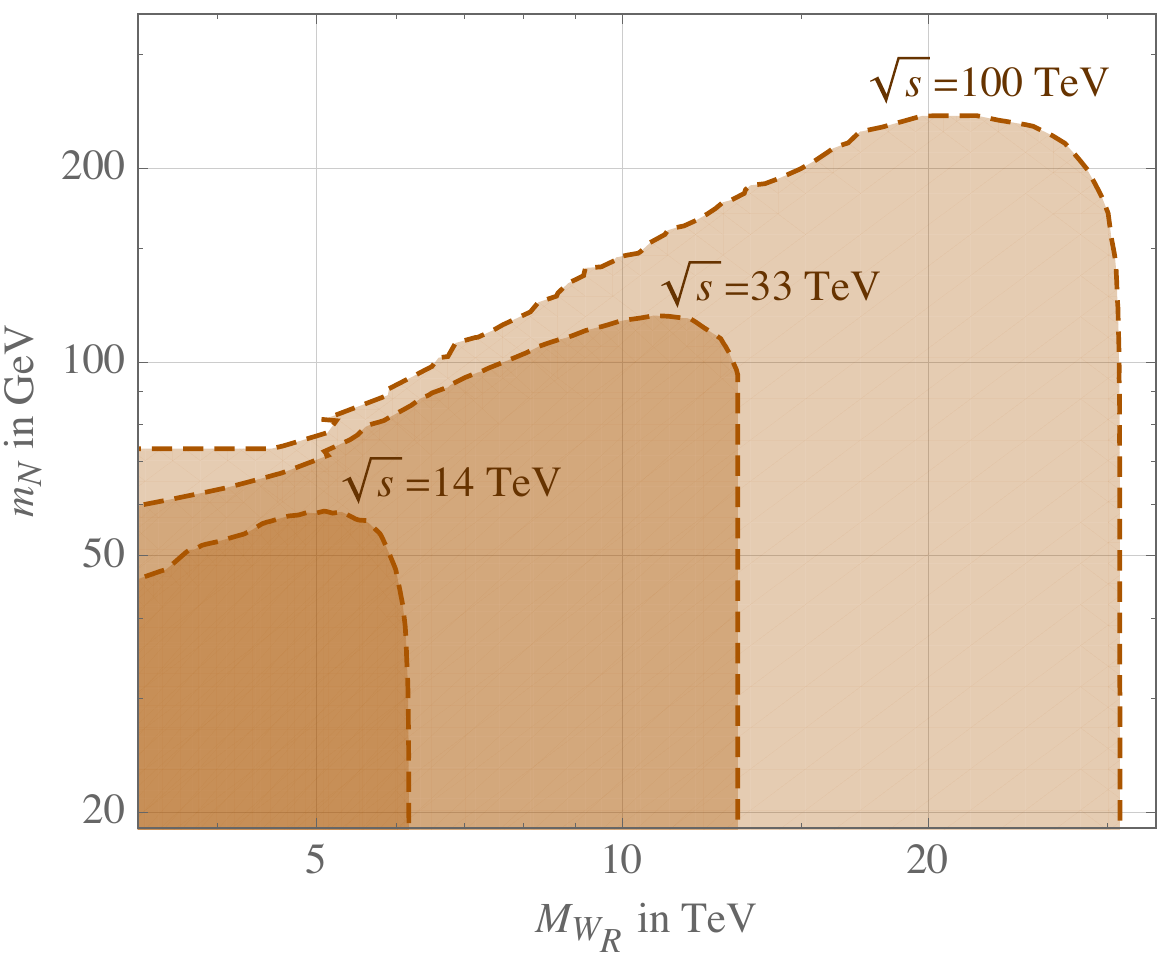}}%
  \caption{The $\ell+$missing energy channel. Left: exclusion region with present LHC data, recast
  from~\cite{ATLASWpMissing}; the average boosted $N$ lifetime is also shown as dashed lines. The
  difference in electron and muon channels is due to the difference in measured events. Right: reach
  of this channel for 14, 33, 100\,TeV center of mass energy; muon and electron channel basically
  coincide.}
\label{fig:lmiss}
\end{figure*}

%
%
\section{The invisible KS} \label{sec:invisible}
\noindent 
A separate assessment can be provided for the region where $N$ decays outside of the detector. In
fact, in this region a very clean signature appears with a high-$p_T$ charged lepton and significant
missing energy carried away by $N$. This happens for fairly light $m_N \lesssim 10\,\GeV$,
which may be motivated by having a warm DM candidate~\cite{LR_WDM}.

The simple $2 \to 2$ kinematics of the process allows for a straightforward recast of the existing $W' \to \ell \nu$ 
searches~\cite{ATLASWpMissing, CMSWpMissing}, as well as sensitivity estimates for future colliders. To this end,
it is useful to compute the distribution over $m_T = 2 p_T$ for signal events with $N$ decaying outside the detector
radius, taken to be $l_0 = 5\,\text{m}$
\begin{equation} \label{eqdsigdpT2Miss}
\begin{split}
  \frac{\dd \sigma}{\dd m_T} = \alpha_2^2 \frac{\pi}{24} p_T \int_{\tau_-}^1 \int_{\frac{\tau_-}{x_1}}^{1} \dd x_{1,2}
  \frac{\left(\hat s - m_N^2 - 2 p_T^2 \right) \pm 1}{\sqrt{\left(\hat s - m_N^2 \right)^2 - 4 p_T^2 \hat s}}
  \\
   \frac{e^{-l_0/L_\pm} \, \varepsilon^\pm_\ell(p_T, \eta_\ell)}{\left(\hat s - M^2 \right)^2 + \left( \Gamma M \right)^2}
   \left |V_{ud} V_{\ell N} \right |^2 f_u(x_{1,2}) f_{\overline d}(x_{2,1}),
\end{split}
\end{equation}
where $\tau_- = \frac1s\big(m_N^2 + 2 p_T^2 + 2 p_T \sqrt{m_N^2 + p_T^2 }\big)$ and $\hat s = x_1
x_2 s$. The sum goes over $u, d$ quarks, both lepton charges and the two $\hat t$ branches
\begin{equation} \label{eqthatpm}
  \hat t_\pm = \frac{\hat s \left( \tau_0 - 1 \right) }{2}\left( 1 \pm \sqrt{ 1 - \frac{4 p_{T}^2}{\hat s (\tau_0 - 1)^2} }\right),
\end{equation}
with $\tau_0 = m_N^2/\hat s$.
The lab frame decay length
\begin{equation}
  L = \frac{p_T}{m_N \, \Gamma_N} \sqrt{1 + \left(1 + \frac{m_N^2}{p_T^2} \right) \sinh(\eta_N)^2},
\end{equation}
is given by $p_T$ and $\exp(2 \eta_N) = -x_1/x_2 \left(1+ \hat s/ \hat t \right)$. The $\varepsilon_\ell$ are
experimentally determined charged lepton efficiency maps usually given in the $p_T-\eta_\ell$ plane, with 
$\exp(2 \eta_\ell) = -x_2/x_1 \left(1+ \hat s/(\hat t - m_N^2) \right)$.

\begin{figure*}[t]
  \centerline{\includegraphics[width=1.6\columnwidth]{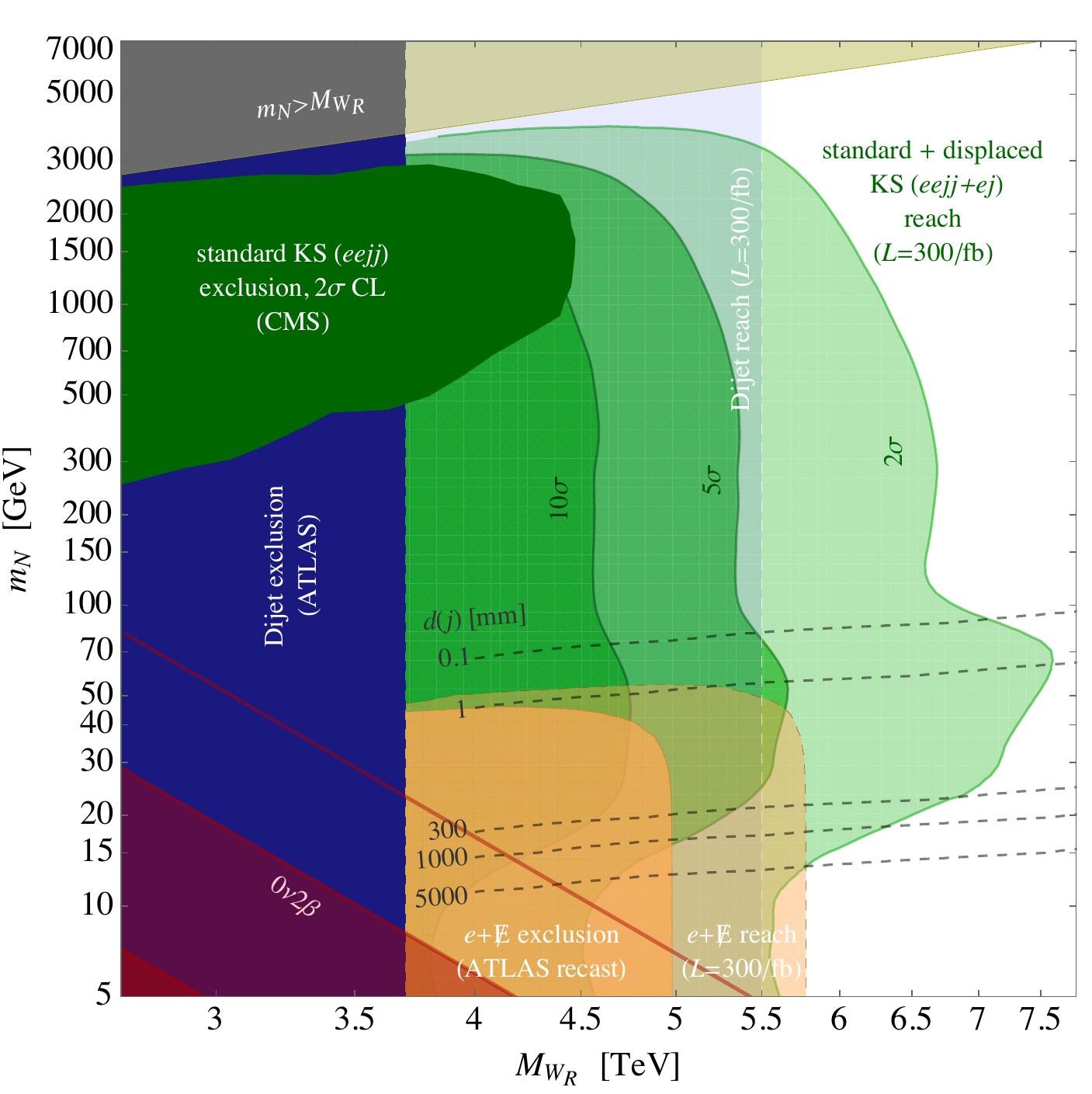}}%
  \caption{Summary plot collecting all searches involving the KS process at LHC, in the electron
    channel. The green shaded areas represent the LH sensitivity to the KS process at $300/$fb,
    according to the present work. The rightmost reaching contour represents the enhancement
    obtained by considering jet displacement.}
  \label{figMasterPlotLHC}
\end{figure*}

The main backgrounds to this process are the SM single $W$, top quark and multi-jet production.
Integrating~\eqref{eqdsigdpT2Miss} in the entire $m_T \in [3-7]\,\TeV$ bin and taking the
corresponding background from~\cite{ATLASWpMissing}, the exclusion in the $M_{W_R}-m_N$ plane is
obtained and shown on the left panel of Fig.~\ref{fig:lmiss} and reproduced below in the
comprehensive Fig.~\ref{figMasterPlotLHC}.

Because of the exponential tail and the boost factor, the limit extends to a very small proper decay
length of $N$ below 1 cm and thus covers the range of $m_N$ well in the $\mathcal O(10\,\GeV$
range for the LHC, as seen in Fig.~\ref{fig:lmiss}. Of course, in the $m_N \to 0$ case, the extremal
limit in~\cite{ATLASWpMissing} is reproduced.

The limits in the electron and muon channels differ due to the difference in the observed data events, not so
much due to the efficiencies or backgrounds. In addition to $e$ and $\mu$, the $\tau$ channel search was also
performed by the CMS collaboration~\cite{CMSWpMissing}. However, because of lower luminosity used in the
search as well as a slightly lower efficiency, the bound goes only up to 3.3 TeV and is not yet competitive with the
di-jet limit.

Due to the cleanliness of the $\ell \slashed E$ final state, the process provides excellent
sensitivity to $W_R$, going almost to the kinematical endpoint of $6.5\,\TeV$ for the HL-LHC program
with $\sqrt s = 14 \text{ TeV}$ and $3 \text{ ab}^{-1}$, see Fig.~\ref{fig:lmiss}. In order to
estimate the sensitivity the background $m_T$ bins were rescaled to proper energies and the global
sensitivity formula in Eq.~\eqref{eq:sensitivity} was used. Assuming the same collected luminosity,
the future $\sqrt s = 33(100)\,\TeV$ $pp$ machines would cover the $W_R$ masses up to $M_{W_R} <
13.5(33)\,\TeV$ and $m_N \lesssim 120(250)\,\GeV$, well in the $\mathcal O(100)\,\GeV$ region,
as seen on the right panel of Fig.~\ref{fig:lmiss}.

%
%
\section{Conclusions and roadmap} \label{sec:conclusions}
\noindent
The case of a TeV-scale Left-Right symmetric extension of the Standard Model, which provides a
complete theory of neutrino masses and an understanding of the origin of parity breaking, still
resists as a viable case, notwithstanding the rapid progress of LHC in probing and excluding the
scales of new physics.
The main channel for discovering the RH gauge boson $W_R$ in connection with the RH neutrino $N$ is
the so called Keung-Senjanovi\'c (KS) process~\cite{Keung:1983uu}, $pp\to W_R\to \ell N\to \ell \ell j j $.
The constraints from direct searches~\cite{CMSKS, ATLASKS}, from flavour changing 
processes~\cite{Bertolini:2014sua, Maiezza:2014ala} and model perturbativity
~\cite{Maiezza:2016bzp} point to a scale of the new RH interaction which is
now at the fringe of the LHC reach, so the residual kinematically accessible range will be probed in
the next year of two.

In this work we reconsidered this process and addressed the regime of light $N$ ($m_N\lesssim
100\,\GeV$) which leads~\cite{Nemevsek:2011hz} to long lived RH neutrino and thus to displaced
vertices from its decay to a lepton and jets. This complements previous studies and gives
a comprehensive overview of the collider reach covering the full parametric space.

To this aim, we classified the signatures resulting from the KS process, depending on the RH
neutrino mass, in four regions: 1) a \emph{standard} region where the final state is $\ell\ell j j
$, with half of the cases featuring same-sign leptons, testifying the lepton number violation.  2)
a \emph{merged} region, with lighter and more boosted $N$, in which its decay products are
typically merged in a single jet including the secondary lepton, resulting in a lepton and a so
called neutrino jet $\ell j_N$. 3) a \emph{displaced} region, for $m_N\sim 10-100\,\GeV$, in which
the merged jet $j_N$ is originated from the $N$ decay at some appreciable displacement from the
primary vertex, typically from mm to 30\,cm where the silicon tracking ends and detection of
displaced tracks becomes unfeasible. 4) an \emph{invisible} region, for $m_N\lesssim 40\,\GeV$, in
which $N$ can decay outside the tracking chambers of even the full detector, leading thus to a
signature of a lepton plus missing energy, $\ell E\!\!\!\!/$.

We assessed the reach in all these regions by scanning the $m_N$, $M_{W_R}$ parameter space, up to
$\mathcal O(10) \text{ TeV}$.  For $W_R$ masses beyond $\sim 5\,\TeV$ the process is dominated by
the off-shell $W_R^*$ production, and we noted that, by this mechanism, for $m_N \lesssim 500\,\GeV$
the final cross section gets an enhancement (see Fig.~\ref{fig:WRproduction}) due to the
typical $W_R^*$ invariant mass $\sim\TeV$ (see Fig.~\ref{fig:WminvLpt}).  This eases probing
the light-$N$ region with respect to previous studies.

The results are summarized in the comprehensive Fig.~\ref{figMasterPlotLHC}.  The analysis of the
novel \emph{displaced} region is offered for the first time in this work and shows that by using the
decay displacement as a discriminating variable this region has a very promising highest potential
of detection, reaching up to $M_{W_R}\simeq 7$--7.5$\,\TeV$.

In order to carry out the above analysis the following procedure was adopted. After noting that
multipurpose event generators do not deal well with long lived particles, we developed a dedicated
generator (see Appendix~\ref{app:kseg} and \cite{lrsite}). This was followed by standard
\textsc{Pythia} hadronization and showering. Also detector simulation had to be updated by
developing custom \textsc{Delphes} modules, in order to realistically detect the jet displacement
(see section~\ref{sec:Regions}). See~\cite{thesis} for additional details.

The basic signature of at least one energetic prompt lepton plus one possibly displaced jet ensures
triggering and allowed us to estimate the relevant background of vector boson(s) plus jets, as
described in section~\ref{sec:Regions}.

The interplay between the primary lepton momentum, jet displacement and other variables calls for an
ad hoc procedure for assessing the LHC sensitivity, whereas standard selection cuts would be
cumbersome and ineffective.  We devised a simple and robust statistical method which generalizes the
$s/\sqrt{s+b}$ measure to binned distributions, and also cross checked it versus the more
sophisticated method using Multi Variate neural networks. The results were broadly consistent but
even better in sensitivity with respect to the neural network approach, which is also much slower.

In Fig.~\ref{figMasterPlotLHC} we report the expected sensitivity in the electron channel as
analyzed in this work, and collect all present constraints. These include the current KS search from
CMS~\cite{CMSKS} and ATLAS~\cite{ATLASKS} and the $W_R \to jj$~\cite{ATLASDijet} excluding up to
$M_{W_R}\lesssim 3.7\,\TeV$. A similar sensitivity on \(Z_{LR}\) from dilepton bounds was reported 
in~\cite{Lindner:2016lpp}, while the \(h, Z' \to NN\) were studied in the context of a related \(B-L\) model
~\cite{Kang:2015uoc, Accomando:2016rpc, Accomando:2017qcs}.
  
In the lower-left part of the plot, we add the region connected with \BB-decay, showing both the parameter space 
excluded by current probes ~\cite{Agostini:2013mzu, KamLAND-Zen:2016pfg} as well as sensitivity of the next
round of experiments. The relevant parameter space coincides by now with the lowest neutrino masses,
i.e.~with the \emph{invisible} region.

The prospects for detection at LHC in the three \emph{standard}, \emph{merged} and \emph{displaced}
regions are put together as the green shaded areas for 2, 5, 10\,$\sigma$ sensitivity. The upper
part of this contour traces the standard KS case of the $\ell\ell j j$ signature, while in the lower
part the displacement helps in raising the sensitivity.

In the intermediate merged region, for which $0.01<m_N/M_{W_R}<0.1$, i.e.\ $m_N\simeq
50$--$500\,\GeV$ before the onset of displacement, we obtain a promising sensitivity to
$M_{W_R}\simeq6.5\,\TeV$, at $2\sigma$ C.L.. This region was analyzed also by the first
study~\cite{Ferrari:2000sp}, reporting a limiting sensitivity to $\sim6\,\TeV$, and also by the
recent work~\cite{Mitra:2016kov} that reported a lower figure, circa 5.2\,\TeV.  Having checked that
the relevant simulated backgrounds are equivalent, we attribute the improvement to our new binning
procedure replacing the usual kinematical cuts. This region is also sensitive to complementary 
searches at the LHeC electron-proton collider with a prompt jet and (possibly
displaced) $ejj$ vertex~\cite{Lindner:2016lxq}.

With an orange area we report the analysis of the \emph{invisible} region, obtained by recasting the
current search for $W'$ in the $\ell E\!\!\!\!/$ signature. It covers the region of $m_N\lesssim
40\,\GeV$, and we can presently exclude up to $M_{W_R}<5\,\TeV$. With $300/$fb of integrated
luminosity LHC will be able to exclude up to circa 5.7\,TeV.
 
The most prominent feature of our results is a sensible improvement of the sensitivity as soon as
the jet displacement is effective as a discriminating variable, see Fig.~\ref{fig_sens_300} for both
muon and electron channels.  For displacements of the order of 10\,mm, one can probe $M_{W_R}$ as
large as $\sim7(7.5)\,\TeV$ in the electron(muon) channel.  For displacements below few mm the
sensitivity could be even larger, as shown by the bump in Fig.~\ref{figMasterPlotLHC} but a
realistic assessment of the vertexing capabilities should be carried out in the concrete detector
environment.

While there are no existing experimental searches that directly address the displaced vertex region,
a very recent study was performed in~\cite{Cottin:2018kmq} by recasting to an existing ATLAS search
for displaced vertices and missing energy~\cite{Aaboud:2017iio}. The authors find the existing
search has poor sensitivity, and propose a relaxed \(N_{trk}\) and \(m_{DV}\) requirements to
significantly enhance the efficiency.  The region of interest for that search is for \(m_N\) below
40 GeV, where the invisible decay proves to be more competitive, see the lower part of
Fig.~\ref{figMasterPlotLHC}.  Nevertheless, an improvement in sensitivity and going below the
fiducial 4\,mm displacement to access higher \(m_N\) seems promising.


From the above results one can conclude that if one extends the current searches by considering also
displacement of jets, in a realistic range up to 30\,cm, the LHC search for the KS process can reach
a sensitivity up to 7--8\,\TeV at 95\%\,C.L., for RH neutrino masses down to $\sim 20\,\GeV$.

Further improvements in the recognition of even more displaced jets, like so called emerging jets,
or displaced muons as distant as the muon chambers are also subject of current
study~\cite{emerging}, and they could extend the sensitivities to even lower RH neutrino masses.


\section{Acknowledgments}

\noindent
The work of MN was supported by the Slovenian Research Agency under the research core funding
No. P1-0035 and in part by the research grant J1-8137.  FN and GP are partially supported by the
H2020 CSA Twinning project No.\ 692194 ``RBI-T-WINNING'' and by the Croatian Science Foundation
(HRZZ) project PhySMaB.

\begin{widetext}
\end{widetext}

\appendix
%
%
\section{Width of $N$} \label{app:nwidth}
\noindent
Computing the three-body decay width becomes involved when masses of decay products have to be taken
into account. In the case of $N$ decaying into a lepton and a quark pair, further complications
arise in the squared amplitude when mass of $N$ becomes comparable with mass of $W_R$, since
the invariant mass of the quark pair cannot be neglected.

However, the width can be computed numerically. Full squared amplitude, although lengthy, is
straightforward to calculate (using FORM, for instance) and phase space can be split into two
pieces: two-body decay of $N$ into lepton and $W_R$, and decay of $W_R$ into a quark pair.  This
introduces a nontrivial integral over invariant mass $q^2$ of a quark pair and over the solid angle
$\dd\Omega^*$ of one of the quarks in the rest frame of $W_R$.  After boosting the quarks into the
rest frame of $N$, the integral over $\dd\Omega$ is simple, since the squared amplitude is a
polynomial in $\cos\theta$.  The width of $N$ for decaying into a lepton of mass $m_l$ and quark
pair with masses $m_1$ and $m_2$ is then
\begin{equation}
\begin{split}
  \frac{\dd\Gamma}{\dd q^2} &= \frac{\alpha_2^2 N_c}{128 \pi} \frac{1}{m_N \left( q^2 - M_{W_R}^2 \right)^2} 
        \\ &
        \times \lambda^{\frac{1}{2}}\left(\frac{m_l^2}{m_N^2},\frac{q^2}{m_N^2}\right)
        \lambda^{\frac{1}{2}}\left(\frac{m_1^2}{q^2},\frac{m_2^2}{q^2}\right) \mathcal{A}(q^2),
\end{split}
\end{equation}
where $\mathcal{A}(q^2)$ is the spin-averaged amplitude with angular dependency integrated out
(coupling constants and scalar part of $W_R$ propagator are pulled out) and
$\lambda(x,y)=1+x^2+y^2-2x-2y-2xy$.  The remaining integral over $q^2$, $(m_1+m_2)^2 \le q^2 \le
(m_N-m_l)^2$, can be easily evaluated numerically to a very high precision.

%
%
\section{$N$ production with off-shell $W_R$} \label{app:offshell}
\noindent 
We collect here the cross section of the KS process via on- and off-shell $W_R$. For ease of
notation the mass and width of $W_R$ are denoted in this section as $M$ and $\Gamma$.

\SEC{On-shell $W_R$ production.}
\begin{align}
  &\hat{\sigma}_{ij}(\hat{s}) = \frac{\alpha_2 \pi^2}{3} \left| V^{\text{\tiny CKM}}_{ij} \right|^2 \delta \left( \hat{s} - M^2 \right),
  \\
  &\sigma = \frac{\alpha_2 \pi^2}{3 s} \sum_{\substack{i=u,c \\ j=\bar{d},\bar{s}}} \left| V^{\text{\tiny CKM}}_{ij} \right|^2
   \int \limits_{\frac{M^2}{s}}^1 \hspace{-0.3em}\frac{\dd x}{x} \, f_{ij}\left( x, \frac{M^2}{x s}, M^2 \right).
\end{align}

\SEC{$N$ production cross section.} The rate for the process
\begin{equation}
  u_i(k_1) + \bar{d}_j(k_2) \to W_R^+ \to l^+(p_1) + N(p_2),
\end{equation}
where $u_i$ is up-type quark and $\bar{d}_j$ is down-type antiquark, at the parton level is
\begin{equation}\label{inv_xsect}
  \frac{\dd\hat{\sigma}_{ij}}{\dd\hat{t}} = \frac{\alpha_2^2 \pi}{12 \hat{s}^2}
  \left| V^{\text{\tiny CKM}}_{ij} \right|^2 \frac{\hat{t}(\hat{t}-m_N^2)}{(\hat{s}-M^2)^2 + M^2\Gamma^2}.
\end{equation}
In the parton CMS frame, $\hat{s} = (\hat{k}_1+\hat{k}_2)^2$, and
\begin{equation}\label{mandelstam_t}
  \hat{t} = (\hat{k}_1-\hat{p}_1)^2 = -\frac{\hat{s}-m_N^2}{2} \left(1 - \cos \theta \right),
\end{equation}
where $\theta$ is the angle between $\hat{\mathbf{k}}_1$ and $\hat{\mathbf{p}}_1$. The total
parton-level cross section is then
\begin{equation}\label{tot_xsection}
  \hat{\sigma}_{ij}(\hat{s}) = \frac{\alpha_2^2 \pi}{72 \hat{s}^2} \left| V^{\text{\tiny CKM}}_{ij} \right|^2
  \frac{(\hat{s}-m_N^2)^2 \left(2 \hat{s} + m_N^2 \right)}{(\hat{s}-M^2)^2 + M^2\Gamma^2}.
\end{equation}
To obtain inclusive rates, convolution with parton distribution is needed,
\begin{align}
  \sigma &= \int \limits_{\frac{m_N^2}{s}}^1 \hspace{-0.5em} \frac{\dd x_1}{x_1}
  \int\limits_{m_N^2}^{x_1 s} \frac{\dd\hat{s}}{s}\,
  \sum_{\substack{i=u,c \\ j=\bar{d},\bar{s}}} \hat{\sigma}_{ij}(\hat{s}) \, f_{ij}\left( x_1, \frac{\hat{s}}{x_1s}; Q^2 \right),
\end{align}
where $\sqrt{s}$ is the center of momentum energy in laboratory frame,
\begin{equation}
\begin{split}
  f_{ij}(x_1,x_2;Q^2) &= f_{i/p}(x_1;Q^2) f_{j/p}(x_2;Q^2)
  \notag \\ &+
  f_{i/p}(x_2;Q^2) f_{j/p}(x_1;Q^2),
\end{split}
\end{equation}
where $f_{i,j/p}(x;Q^2)$ are parton distribution functions evaluated at momentum fraction $x$ and
factorization scale $Q^2 = \hat{s}$ (default in \textsc{\textsc{MadGraph}} for KS process). Difference
between production of $W_R^+$ and $W_R^-$ is only in the parton distributions.

Relevant (kinematical) distributions can easily be derived from \eqref{inv_xsect} and
\eqref{tot_xsection} by inserting the appropriate $\delta$-functions, for instance
\begin{equation}
  \frac{\dd\hat{\sigma}}{\dd y} = \int \frac{\dd\hat{\sigma}}{\dd\hat{t}}\,
  \delta \left( y - g(\hat{t}) \right) \, \dd \hat{t}.
\end{equation}

\SEC{$W_R$ invariant mass distribution.}
Invariant mass distribution for $W_R$ is simply
\begin{equation}
  \frac{\dd\hat{\sigma}_{ij}}{\dd M^2} = \hat{\sigma}_{ij}(\hat{s}) \, \delta(M^2-\hat{s})
\end{equation}
and then
\begin{align}
  \frac{\dd\sigma}{\dd M^2} &= \sum_{\substack{i=u,c \\ j=\bar{d},\bar{s}}} \hat{\sigma}_{ij}(M^2)
  \int\limits_{\frac{M^2}{s}}^1 \hspace{-0.3em} \frac{\dd x_1}{x_1s}\, f_{ij}(x_1, \frac{M^2}{x_1 s}; Q^2).
\end{align}

\newcommand{\pt}{p_{{}_T}}
\SEC{Prompt lepton $\pt$ distribution.}
Transverse momentum distribution for the prompt lepton is obtained by inserting
\begin{equation}
  1 = \int\limits_{0}^{|\mathbf{p}_1|} \dd \pt \delta(\pt - |\mathbf{p}_1|\sin\theta)
\end{equation}
into \eqref{inv_xsect} and integrating over $\hat{t}$,
\begin{equation}
\begin{split}
  \frac{\dd\hat{\sigma}_{ij}}{\dd \pt} &= \frac{\alpha_2^2 \pi}{6} \left| V^{\text{\tiny CKM}}_{ij} \right|^2
  \frac{\pt}{\sqrt{(\hat{s}-m_N^2)^2-4\hat{s}\pt^2}}
  \\ &\times
  \frac{\hat{s}-2\pt^2-m_N^2}
  {(\hat{s}-M^2)^2 + M^2\Gamma^2}.
\end{split}
\end{equation}
The convolution with parton distributions gives then
\begin{equation}
\begin{split}
  \frac{\dd\sigma}{\dd\pt} &= \int\limits_{\frac{\varepsilon_T^2}{s}}^1
  \frac{\dd x_1}{x_1} \int\limits_{\varepsilon_T^2}^{x_1s} \frac{\dd\hat{s}}{s}\,
  \sum_{\substack{i=u,c \\ j=\bar{d},\bar{s}}} \frac{\dd\hat{\sigma}_{ij}}{\dd\pt}
  f_{ij}(x_1, \frac{\varepsilon_{{}_T}^2}{x_1 s}; Q^2),
\end{split}
\end{equation}
where $\varepsilon_{{}_T} = \pt + \sqrt{\pt^2+m_N^2}$.

\section{Generation of events for small $N$ width}

\label{app:kseg}
\noindent
The cross section for the full KS process
\begin{equation}
  q_i(k_1) \bar{q}_j(k_2) \to l^{\pm}(p_1) l^{\pm}(p_2) j(p_3) j(p_4)
\end{equation}
can be written as
\begin{equation}
  \sigma = \int\dd x_1 \dd x_2 \sum_{u,d,h} \hat{\sigma}_{ud,h}(x_1, x_2) f_{ud}(x_1, x_2, Q^2),
 \end{equation}
where $\hat{\sigma}_{ud,h}$ is the partonic cross section with quark flavors $u$ and $d$ and helicity configuration
denoted by $h$. The phase space in $\hat{\sigma}_{ud,h}$ can easily be split into a sequence of 2-particle ones, for example
\begin{equation}
\begin{split} \label{ps_mapping}
 \dd \Phi(k_1+k_2 \to \textstyle \sum_{1}^4 p_i) = \dd\Phi(k_1+k_2\to p_1+q_{234}) 
 \\
 \dd\Phi(q_{234}\to p_2+q_{34}) \dd\Phi(q_{34}\to p_3+p_4) \frac{\dd q_{234}^2 \dd q_{34}^2}{(2\pi)^2} ,
\end{split}
\end{equation}
and each of them is simply
\begin{equation}
  \dd\Phi(P\to p_1+p_2) = \frac{1}{8\pi} \lambda^{\frac{1}{2}} \left(\frac{p_1^2}{P^2}, 
  \frac{p_2^2}{P^2}\right) \frac{\dd\Omega}{4\pi},
\end{equation}
where $\dd\Omega = \dd\phi\,\dd\!\cos\theta$ is the solid angle of $\mathbf{p}_1$ or $\mathbf{p}_2$
in the rest frame of $P$ with respect to some axis, most conveniently taken in the direction of
$\mathbf{P}$.  In order to generate the events, angles and invariant masses in~\eqref{ps_mapping},
as well as parton momentum fractions, $x_1$ and $x_2$ are randomly sampled. Eq.~\eqref{ps_mapping}
corresponds to one possible phase space mapping, given by the kinematical structure of a diagram(s)
describing the process.
	
Difficulties in Monte Carlo event generation of the KS process arise from the sharp (and dominant)
peak in the invariant amplitude coming from a very small width in the neutrino propagator. Adaptive
integration methods may not be able to handle such extreme cases, however this problem can be easily
solved by sampling the appropriate phase space variables according to the Breit-Wigner distribution
(importance sampling).

Since KS process consists of multiple subprocesses (helicity combinations, ingoing and outgoing
quarks) each with one diagram for opposite sign leptons or two diagrams for same sign leptons in
the final state, events are generated using the multichannel method.  Each channel corresponds to a
specific subprocess and phase space mapping for different diagrams and carries a weight $\alpha_i$
and a probability density $g_i(\mathbf{x})$, such that $g(\mathbf{x}) = \sum_i \alpha_i
g_i(\mathbf{x})$ and $\sum_i\alpha_i=1$, where $\mathbf{x}$ are phase space variables. Weights
$\alpha_i$ are thus probabilities of selecting different channels and can be optimized during event
generation fo better performance~\cite{weight_optim}. A suitable way to optimize $\alpha_i$ was
proposed in~\cite{madevent}, by introducing a basis of functions
 \begin{equation}
  f(\mathbf{x}) = \sum_i f_i(\mathbf{x}), 
  \quad f_i(\mathbf{x}) = \frac{ \left|\mathcal{M}_i \right|^2}{
  \sum_j \left| \mathcal{M}_j \right|^2} \left| \mathcal{M}_{\text{tot}} \right|^2,
\end{equation}
where $\mathcal{M}_{\text{tot}}=\sum_i\mathcal{M}_i$. The integral is now the sum of contributions with different
peaking structures (contained in the amplitudes $\mathcal{M}_i$),
\begin{equation}
  I = \int\dd\mathbf{x}\,f(\mathbf{x}) = \sum_i \int\dd\mathbf{x}\,g_i(\mathbf{x})
  \frac{f_i(\mathbf{x})}{g_i(\mathbf{x})} = \sum_i I_i,
\end{equation}
and optimized weights are $\alpha_i = I_i/I$. This approach avoids the evaluation of all
$g_i(\mathbf{x})$ for every point in phase space and the complications related to the correlations
between $\alpha_i$ when the number of channels is large.

For the event generation software, as well as custom detector simulation and analysis, visit te web
site~\cite{lrsite}.

%
%
%
\section{Assessing sensitivity} \label{app:sensitivity}
\noindent
It is a common problem, prior to having experimental data available, to assess the sensitivity of an
experiment to a given hypothesis of new physics, defined as the number of signal ($s$) events
expected on top of a number of background ($b$) events.  
These may be single numbers as in a simple counting experiment, or binned distributions in relevant
kinematical variables like in the present case.

In a (Poisson) counting experiment, equivalent to the case of a single bin, it is customary to
define the sensitivity as $s/\sqrt{s+b}$. This can be understood as a measure of the ``separation''
between the expected distributions in the hypotheses of background-only and background plus signal
(see below).

In the case of more bins distributed in one or more kinematical variables, the usual procedure is to
define cuts that exclude regions in which backgrounds dominate, and finally assess the surviving
number of signal and background events ($S_{tot}$, $B_{tot}$). The choice of cuts must be optimized
in order to maximize the global sensitivity e.g.~$S_{tot}/\sqrt{S_{tot}+B_{tot}}$.  This procedure
can become quite complex with an increasing number of variables and if the region that one would
like to cut has a nontrivial shape in their multidimensional space. Sometimes the procedure of
cutting away the high-background low-sensitivity bins is even impossible.

Consequently, one can ask whether one could just define a measure that automatically weighs the
various bins according to their contribution to the sensitivity. The answer is simple and amounts to
adding in quadrature the sensitivities associated to each bin, such that the global sensitivity is defined as (\ref{eq:sensitivity})
\begin{equation} \label{eq:sensitivityapp}
  \text{sensitivity}\ \Sigma=\left[\sum_{i\in\text{bins}}\frac{s_i^2}{s_i+b_i}\right]^{1/2},
\end{equation}
where $s_i$, $b_i$ are the expected number of signal and background events in each bin and we stress
that the sum runs on the full grid of bins in the multidimensional space of kinematical variables.
The resulting method is able to assess the global sensitivity of the experiment in a straightforward
manner without having to impose cuts.

We discuss here first the formal justification, then the statistical uncertainty on this measure, as
well as the systematics due to different binning.

For a Poisson counting experiment with expected number of events $\mu s+b$, the likelihood function is
\begin{equation}
  L(\mu) = \frac{(\mu s+b)^n}{n!}e^{-(\mu s+b)},
\end{equation}
where $s$ ($b$) is the number of signal (background) events and $\mu$ is the signal rate parameter,
i.e.\ $\mu=0$ corresponds to the background only hypothesis, while $\mu=1$ to the signal plus
background hypothesis. The maximum likelihood estimator of $\mu$ is $\hat{\mu} = (n-b)/s$ and has 
clearly expectation $E[\hat\mu]=\mu$, while its variance is
\begin{equation}
  V[\hat{\mu}] = E[\hat{\mu}^2]-E[\hat{\mu}]^2 = \frac{\mu s+b}{s^2}\,.
\end{equation}
At $\mu=1$ (signal hypothesis) the standard deviation of the estimator $\hat{\mu}$ is
$\sigma_{\hat{\mu}}=\sqrt{s+b}/s$ and thus $s/\sqrt{s+b}$ can be interpreted as the expected
significance with which one could reject $s$ if the signal is absent~\cite{statrpp}.

One can proceed similarly in the case of more bins, but it is useful to first rescale $\mu$ into
$\nu=\mu\,s/\sqrt{s+b}$ such that the likelihood is
\begin{equation}
  L(\nu) = \frac{(\nu\sqrt{s+b}+b)^n}{n!}e^{-(\nu\sqrt{s+b}+b)}
\end{equation}
and the estimator is $\hat{\nu} = (n-b)/\sqrt{s+b}$. This has clearly expectation $\nu = s \mu
/\sqrt{s+b}$ and variance $(\nu \sqrt{s+b} + b)/(s+b)$. In the hypothesis of signal, expectation is
$s/\sqrt{s+b}$ and variance is 1.

Now we consider together all (uncorrelated) bins. In the case of signal the distribution of the vector
$\{\hat\nu_i\}$ is centered at position $\{s_i/\sqrt{s_i+b_i}\}$, still with unit variance 1 in each
dimension. So, the distribution of $\{\hat\nu_i\}$ in the case of signal is peaked there, inside
a ``hypersphere'' of radius 1. On the other hand, the case of no signal is represented by the
origin, $\{\hat\nu_i=0\}$.

Thus, the definition of sensitivity in~(\ref{eq:sensitivity}) represents the distance of the origin
from the center of the unit hypersphere, and it can indeed be taken as a measure of the significance
with which one can exclude the signal in case of no signal. The sum in quadrature
in~(\ref{eq:sensitivityapp}) takes contributions from the bins where significance is high, and
negligible increase from the bins with no signal or dominant background, as it has to be.

\SEC{Uncertainty in sensitivity.} Let us briefly discuss the statistical and systematic uncertainties which affect the sensitivity measure~(\ref{eq:sensitivity}).

We can discuss the statistical uncertainty, if the distribution in bins remains smooth as binning is
refined, i.e.\ if locally $s$, $b\sim 1/N_{bins}$.  In this case, for each bin the uncertainty on its
contribution to the sensitivity, $s_i^2/(s_i+b_i)$, is $[s_i^3 (4 b_i + s_i)/(s_i + b_i)^3]^{1/2}$, such
that the uncertainty on $\Sigma$ is obtained by summing in quadrature all bins and is
\begin{equation}\label{eq:stat}
  \sigma_{stat} \left( \Sigma \right) = \frac{1}{2\Sigma}\left[\sum_i \frac{s_i^3 \left(4 b_i + s_i \right)}{
  (s_i + b_i)^3} \right]^{\frac12}
\end{equation}
Notice that all terms in the sums in~(\ref{eq:sensitivity}) and (\ref{eq:stat}) scale as
$\sim1/N_{bins}$, so the final statistical uncertainty is not increased with finer binning.

More interesting is the systematic error that can arise when, in refining the binning, one hits the
limit of smoothness of the distribution. Typically this happens first for the background, that may
be simulated with less statistics, due to the higher required computing time.  One can ask what
happens in case in some region of parameter space this overbinning leads to a background events 
concentrated in isolated bins, while the signal is still smooth. In this case, the contribution to
the sum (\ref{eq:sensitivity})\ will contain an increasingly larger number of bins with just signal,
increasing the sensitivity, plus a fixed number of bins with background.  As a limiting example, let
us describe the case in which in a region with total events $S$ and $B$, all background is
concentrated in a single isolated bin, while the signal still scales as $1/N_{bins}$. For simplicity
we assume also that in this region the signal distribution is constant, $s_i=S/N_{bin}$. In the
limit of very fine binning the isolated background bin disappears from the final result of the sensitivity:
\be
\Sigma_1=\left[A + S \right]^{1/2},
\ee
where $A$ represent the contribution of the rest of the bin.  This should be compared with the
standard smooth background case: 
\be
\Sigma=\left[A+\frac{S^2}{S+B}\right]^{\frac12}. 
\ee
The difference between these two is an estimate of the systematic error induced by overbinning the background, and
it can be approximated as (for non negligible $\Sigma$):
\be
  \sigma_{syst}\left(\Sigma\right)\simeq\frac1{2\Sigma} \frac{B S}{B+S}\simeq \frac1{2\Sigma}\min(B,S)\,.
  \label{eq:systgen}
\ee

From this result one finds that the systematic error can be quite small even with large $B$: indeed,
if $S$ is small in the regions where there is isolated background $B$, there is small contribution
to the sensitivity and also to the uncertainty.

Similar to what is done in the MVA analysis (see e.g.\ \cite{MVA}) the optimal approach would be to
estimate the magnitude of $S$ by exploring a region around isolated backgrounds, in order to check
whether an increased contribution to the uncertainty is indeed present or not, and wether a more
coarse grained binning would be needed.  To delimit the regions where the background has isolated
bins is however a typically hard task, and thus it is difficult to asses the relevant $S$.
Fortunately, from (\ref{eq:systgen}) we note that it is actually sufficient to limit the value of
$B$, i.e.\ of the total number of background events that remain in isolated bins. In this way one
can control the systematic uncertainty, albeit overestimating it. The consequent upper limit is the
figure of merit which we quote in the table in the text,
\be
\sigma_{syst}\left(\Sigma\right) < B_{isolated}/{2\Sigma}\,,
\label{eq:syst}
\ee
in order to make sure that overbinning of background has negligible impact on our results.

\def\arxiv#1[#2]{\href{http://arxiv.org/abs/#1}{[#2]}}
\def\Arxiv#1[#2]{\href{http://arxiv.org/abs/#1}{#2}}

\end{document}